\documentclass[aps,prd,twocolumn,groupedaddress,showpacs,nofootinbib]{revtex4-1}

\usepackage{graphicx}
\usepackage{amsmath}
\usepackage{amssymb}
\usepackage{aas_macros}
\usepackage{color}
\usepackage{comment}

\newcommand{\mch}{\ensuremath{\mathrm{M}_\mathrm{Ch}}}
\newcommand{\msun}{\ensuremath{\mathrm{M}_\odot}}
\newcommand{\ia}{SN~Ia}
\newcommand{\iae}{SNe~Ia}

\newcommand{\liae}{Type~Ia supernovae}
\newcommand{\plotsize}{1.0}

\begin{document}
\title{Neutrino and gravitational wave signal of a
  delayed-detonation model of Type Ia supernovae}
\author{Ivo R. Seitenzahl}
\email{ivo.seitenzahl@anu.edu.au}
\affiliation{Research School of Astronomy and Astrophysics, The Australian
  National University, Cotter Road, Weston Creek, ACT, 2611, Australia}
\affiliation{ARC Centre of Excellence for All-Sky Astrophysics (CAASTRO)}
\affiliation{Institut f{\"u}r Theoretische Physik und Astrophysik,
  Universit\"at W\"urzburg, Emil-Fischer-Str.~31, D-97074 W{\"u}rzburg, Germany}
\affiliation{Max-Planck-Institut f{\"u}r Astrophysik,
  Karl-Schwarzschild-Str.~1, D-85741 Garching, Germany}
\author{Matthias Herzog} 
\affiliation{Max-Planck-Institut f{\"u}r Astrophysik,
  Karl-Schwarzschild-Str.~1, D-85741 Garching, Germany}
\author{Ashley J. Ruiter} 
\affiliation{Research School of Astronomy and Astrophysics, The Australian
  National University, Cotter Road, Weston Creek, ACT, 2611, Australia}
\affiliation{ARC Centre of Excellence for All-Sky Astrophysics (CAASTRO)}
\affiliation{Max-Planck-Institut f{\"u}r Astrophysik,
  Karl-Schwarzschild-Str.~1, D-85741 Garching, Germany}
\author{Kai Marquardt}
\affiliation{Zentrum f{\"u}r Astronomie der Universit{\"a}t Heidelberg, 
       Institut f{\"u}r Theoretische Astrophysik, Philosophenweg 12, 
       D-69120 Heidelberg, Germany}
\affiliation{Institut f{\"u}r Theoretische Physik und Astrophysik,
  Universit\"at W\"urzburg, Emil-Fischer-Str.~31, D-97074 W{\"u}rzburg, Germany}
\affiliation{Heidelberger Institut f\"{u}r Theoretische Studien,
  Schloss-Wolfsbrunnenweg 35, 69118 Heidelberg, Germany}
\author{Sebastian T.~Ohlmann}
\affiliation{Zentrum f{\"u}r Astronomie der Universit{\"a}t Heidelberg, 
       Institut f{\"u}r Theoretische Astrophysik, Philosophenweg 12, 
       D-69120 Heidelberg, Germany}
\affiliation{Institut f{\"u}r Theoretische Physik und Astrophysik,
  Universit\"at W\"urzburg, Emil-Fischer-Str.~31, D-97074 W{\"u}rzburg, Germany}
\affiliation{Heidelberger Institut f\"{u}r Theoretische Studien,
  Schloss-Wolfsbrunnenweg 35, 69118 Heidelberg, Germany}
\author{Friedrich K. R\"opke} 
\affiliation{Zentrum f{\"u}r Astronomie der Universit{\"a}t Heidelberg, 
       Institut f{\"u}r Theoretische Astrophysik, Philosophenweg 12, 
       D-69120 Heidelberg, Germany}
\affiliation{Institut f{\"u}r Theoretische Physik und Astrophysik,
  Universit\"at W\"urzburg, Emil-Fischer-Str.~31, D-97074 W{\"u}rzburg, Germany}
\affiliation{Heidelberger Institut f\"{u}r Theoretische Studien,
  Schloss-Wolfsbrunnenweg 35, 69118 Heidelberg, Germany}
\bibliographystyle{apsrev}

\date{\today}


\begin{abstract}
The progenitor system(s) and the explosion mechanism(s) of Type Ia supernovae
(SNe Ia) are still under debate. Non-electromagnetic observables, in particular
gravitational waves and neutrino emission, of thermonuclear supernovae are a
complementary window to light curves and spectra for studying these enigmatic
objects. A leading model for SNe~Ia is the thermonuclear incineration of a
near-Chandrasekhar mass carbon-oxygen white dwarf star in a
``delayed-detonation''. We calculate a three-dimensional hydrodynamic explosion
for the N100 delayed-detonation model extensively discussed in the literature,
taking the dynamical effects of neutrino emission from all important contributing
source terms into account. Although neutrinos carry away
$2\times10^{49}\,\mathrm{erg}$ of energy, we confirm the common view that
neutrino energy losses are dynamically not very important, resulting in only a
modest reduction of final kinetic energy by two per cent.
We then calculate the gravitational wave signal from the time evolution of the
quadrupole moment. Our model radiates $7\times10^{39}\,\mathrm{erg}$ in
gravitational waves and the spectrum has a pronounced peak around
$0.4\,\mathrm{Hz}$. Depending on viewing angle and polarization, we find that
the future space-based gravitational wave missions DECIGO and BBO would be able
to detect our source to a distance of ${\sim}1.3\,\mathrm{Mpc}$. We predict a
clear signature of the deflagration-to-detonation transition in the neutrino and
the gravitational wave signals. If observed, such a feature would be a strong
indicator of the realization of delayed-detonations in near-Chandrasekhar mass
white dwarfs.
\end{abstract}
\maketitle

\section{Introduction}
\label{intro}
The nature of the progenitor system(s) and the explosion mechanism(s) of 
\liae\ (\iae) are still under debate. 
Although synthetic observables such as time dependent spectra or light
curves have been calculated for numerous explosion models
\citep[e.g.][]{kasen2009a,kromer2010a,pakmor2012a,kromer2013a,sim2013a},
efforts aimed at unambiguously identifying the explosion scenario
based on a comparison of the model spectral evolution to observations
remain inconclusive \citep{roepke2012a}. The same holds for
the \emph{reverse} process of inferring the composition profile of the
supernova using the techique of abundance tomography
\citep{stehle2005a,mazzali2008a,hachinger2009a,hachinger2013a}.
Complementary approaches aimed at distinguishing the nature of the progenitor
systems include the search for the contribution to the early-time light curve
from a possible companion \citep{kasen2010a, bloom2012a, brown2012a},
Na I D line absorption features interpreted as evidence for circumstellar
material \citep{patat2007a,sternberg2011a,dilday2012a}, late-time bolometric
light curves \citep{seitenzahl2009d,roepke2012a,kerzendorf2014a}, gamma-ray
emission \citep{sim2008a, maeda2012b, summa2013a}, setting upper limits
on the single degenerate scenario in old stellar populations from the lack
of X-ray flux associated with the accretion \citep{gilfanov2010a}, comparing
predictions of SN~Ia progenitor models with the properties of observed
SN remnants \citep{badenes2007a,yamaguchi2015a}, searching for the companion
stars around young supernova remnants
\citep{ruiz-lapuente2004a,kerzendorf2009a,schaefer2012a},
or constraining the circumstellar medium and hence the evolutionary channel from
radio observations \citep{horesh2012a,chomiuk2012a}. Another approach aims at
constraining the scenarios via rates and delay times, either by comparing
prediction of binary population synthesis to observations \citep{yungelson2000a,
  ruiter2009a,ruiter2011a,toonen2012a,ruiter2013a,mennekens2010a,wang2012a} 
or by inferring e.g. merger rates from the statistics of observed WD binary
systems \citep{badenes2012a}. 

In this work, we determine and discuss the neutrino- and the gravitational wave
signal of the N100 explosion model of \citet{seitenzahl2013a}.
N100 is a recent, state-of-the-art, three-dimensional hydrodynamic explosion 
model of a delayed-detonation in a near Chandrasekhar-mass (\mch)
carbon-oxygen (CO) white dwarf (WD). Such delayed-detonation models 
are among the most promising candidates to explain a substantial fraction of
the \ia\ population \citep[e.g.][]{hillebrandt2000a}.

We have chosen the N100 model for our study
since its synthetic light curves and spectra make it a 
rather promising candidate for a ``normal'' \ia\ \citep{sim2013a}.
Furthermore, the nucleosynthetic yields and electromagnetic emission of the N100
model have already been used as representative of near-\mch\ delayed-detonations
in the single degenerate scenario in several recent publications
\citep{roepke2012a,summa2013a,liu2013a,seitenzahl2013b,seitenzahl2015a}.
We emphasize that we actually re-calculate the N100 model from
\citet{seitenzahl2013a}, this time also taking internal energy losses due to
neutrino escape from all contributing source terms dynamically into account. 
For completeness, we therefore refer to this model hereafter as N100${\nu}$.

\section{Neutrino signal}
\label{sec:nu}
The maximum central density of WDs is 
$\rho_c^\mathrm{max}\lesssim1\times10^{10}\,\mathrm{g\,cm^{-3}}$
\citep[see e.g.][]{timmes1992a}. Neutrino trapping starts only at densities
of $\sim 10^{11}$ to $10^{12}\,\mathrm{g\,cm^{-3}}$ \citep{bethe1990a}, WD matter
is therefore always transparent to neutrinos; in good approximation,
they leave the star without any interaction. This simplifies the
neutrino physics greatly, because we can assume that the energy that
the neutrinos get when they are created is immediately lost to the
star. Nevertheless, in the hot ashes of burnt CO matter, the thermodynamic
conditions are such that copious amounts of neutrinos can be emitted.
Calculations of the neutrino signal for a few SN Ia explosion models have been
presented in the literature.
\citet{nomoto1984a} and \citet{kunugise2007a} calculated the neutrino signal for
the W7 model \citep{nomoto1984a}, a one-dimensional fast deflagration of a
${\sim}1.37\, \msun$ CO WD. \citet{odrzywolek2011a} calculated the neutrino
signal for the n7d1r10t15c model \citep{plewa2007a}, a two-dimensional pure
deflagration of a ${\sim}1.36\, \msun\ $ CO WD and the Y12 model
\citep{plewa2007a}, a two-dimensional ``detonating-failed-deflagration''
(gravitationally confined detonation) model of
the same 1.36 \msun\ CO WD. Although \citet{odrzywolek2011a} refer to the Y12
model as a delayed-detonation, this term is usually reserved for SNe Ia that
undergo a deflagration-to-detonation transition (DDT) when the deflagration
flame front reaches lower densities where it gets shredded by turbulence and
the mixing of cold fuel and hot ash triggers a detonation \citep{khokhlov1991a}.
The explosion mechanism in the Y12 model is fundamentally different in that the
detonation is triggered not during the rising phase of the deflagration, but
rather at a much later time when deflagration ashes that are gravitationally
bound to the surface of the WD collide, see
\citet[e.g.][]{plewa2004a,seitenzahl2009c}.

Previously considered multi-dimensional SN Ia models represent only peculiar
events, i.e.\ pure turbulent deflagration and gravitationally confined
detonation models. Here, we present neutrino luminosities for a
three-dimensional delayed-detonation model of a ${\sim}1.40\,\msun\ $ CO~WD
that underwent a DDT.
These explosion models have been extensively studied in the
literature, and may be able to explain ``normal'' SNe Ia
\citep[e.g.][]{sim2013a,hillebrandt2013a}. For details about our implementation
of the DDT model see \citet{ciaraldi2013a}. We distinguish between \emph{weak}
neutrinos produced in nuclear reactions and \emph{thermal} neutrinos produced
by thermal plasma processes. In the following two sections, we introduce the
physical processes that generate the neutrinos during thermonuclear
burning in WDs.

\subsection{Weak neutrinos}
\label{sec:weaknu}
\emph{Weak neutrinos} are emitted in nuclear reactions involving the
weak interaction\footnote{Of course, also ``thermal'' neutrinos are
  created due to the weak interaction, but their production does not
  involve nuclear reactions.}; the amount of emitted energy
depends on specific reaction rates and therefore on the composition of
the matter. The reaction rates themselves depend strongly on
temperature and density. Weak neutrinos and antineutrinos are produced
in electron captures
\begin{align}
p+e^-&\rightarrow n+\nu_e\label{eq:ecapture},
\end{align}
positron capture processes
\begin{align}
n+e^+&\rightarrow p+\bar{\nu}_e\label{eq:poscapture},
\end{align}
$\beta+$ decays
\begin{align}
p&\rightarrow n+e^++\nu_e\label{eq:beta+},
\end{align}
and $\beta-$ decays
\begin{align}
n&\rightarrow p+e^-+\bar{\nu}_e\label{eq:beta-}.
\end{align}
All reactions, which can also occur inside nuclei, change the electron
fraction $Y_e$ (defined as the number of electrons per baryon) of the matter.

Neutrinos from electron captures (\ref{eq:ecapture}) are
abundantly produced in the neutronization processes that take place
during and immediately after thermonuclear burning.
As we will see, the energy released
in reactions~(\ref{eq:ecapture}--\ref{eq:beta-}) dominates the
energy released in all other neutrino generating processes in the
physical environment of thermonuclear supernovae.

In particular, electron capture reactions (\ref{eq:ecapture}) play an
important role during the thermonuclear burning of WD matter.
They lead to neutronization, expressed as decreasing $Y_e$,
lowering the electron degeneracy pressure and thereby affecting the dynamics
of the fluid.

Electron captures on the constituent parts of CO WDs,
$^{12}\mathrm{C}$ and $^{16}\mathrm{O}$, would occur via the following
reactions \citep{shapiro1983a}:
\begin{align}
  ^{12}\mathrm{C}+e^-&\rightarrow {^{12}\mathrm{B}+e^-}
  \rightarrow {^{12}\mathrm{Be}}\\
  ^{16}\mathrm{O}+e^-&\rightarrow {^{16}\mathrm{N}+e^-}
  \rightarrow {^{16}\mathrm{C}}.
\end{align}
The threshold energy for the first reaction chain is
$13.37\,\mathrm{MeV}$, for the second reaction chain
$10.42\,\mathrm{MeV}$. The Fermi energies of the electrons match these
values at densities of
$\rho_\mathrm{ec,C}=3.9\times10^{10}\,\mathrm{g\,cm^{-3}}$ and
$\rho_\mathrm{ec,O}=1.9\times10^{10}\,\mathrm{g\,cm^{-3}}$,
respectively \citep{shapiro1983a}. 

As mentioned above, central densities of accreting CO WDs
are not expected to exceed 
$\rho_c\sim9\times10^{9}\,\mathrm{g\,cm^{-3}}$ \citep{timmes1992a,woosley1997b},
well below $\rho_\mathrm{ec,C}$ and $\rho_\mathrm{ec,O}$, before thermonuclear
burning starts in the central region.
Therefore, for densities appropriate for CO WDs, no electron
captures on the most abundant unburnt nuclei occur. They are only possible after
the start of thermonuclear burning, in particular on the hot ``ashes''.

Using the weak interaction rates of \citet{langanke2001a},
\citet{seitenzahl2009a} calculated extensive tables of
the time rate of change of the specific energy $\dot{E_\nu}$ and the time rate
of change of the electron fraction $\dot{Y_e}$ in these ashes for a three
dimensional grid in density, temperature, and $Y_e$. We have extended their tables to
determine the mean electron neutrino and electron antineutrino energies by implementing
equation 17 of \citet{langanke2001a} into the nuclear statistical equilibrium code of
\citet{seitenzahl2009a}. We interpolate in the tables to determine the amount of energy 
carried away by weak neutrinos for each grid cell and simply subtract it from the 
internal energy of the cell in every time step. The mean neutrino energies are determined
by post-processing the output of the hydrodynamics simulations, once again interpolating in the 
tables, utilizing the NSE mass fraction, electron fraction, temperature, and density stored for each cell.

\subsection{Thermal neutrinos}\label{sec:thermalnu}
In the hot plasmas of stellar interiors, neutrinos are also emitted
due to processes that are independent of nuclear reactions; the amount
of emitted energy depends only on the density and temperature of the
matter. Most important are neutrinos originating from
electron-positron pair annihilation, photoemission, plasmon decay,
bremsstrahlung, and recombination processes (for a brief description
and summary of the individual contributions of these processes see
\citep{itoh1996a}).  Neutrinos from these processes are often combined
as \emph{thermal neutrinos}, they comprise neutrinos and antineutrinos of all
flavors, whereas in reactions~(\ref{eq:ecapture}--\ref{eq:beta-})
only electron and anti-electron neutrinos are produced.

We calculate the energy that is released from the WD due to
the emission of \emph{thermal neutrinos} using a 
code\footnote{\url{http://cococubed.asu.edu/code_pages/nuloss.shtml}} 
in which analytic fitting formulas from \citet{itoh1996a} are implemented. 
In every time step we simply subtract the resulting locally calculated
neutrino energies from the internal energy of each cell.

\subsection{Results and Discussion}
\begin{figure}[h]
  \includegraphics[width=\plotsize\columnwidth]{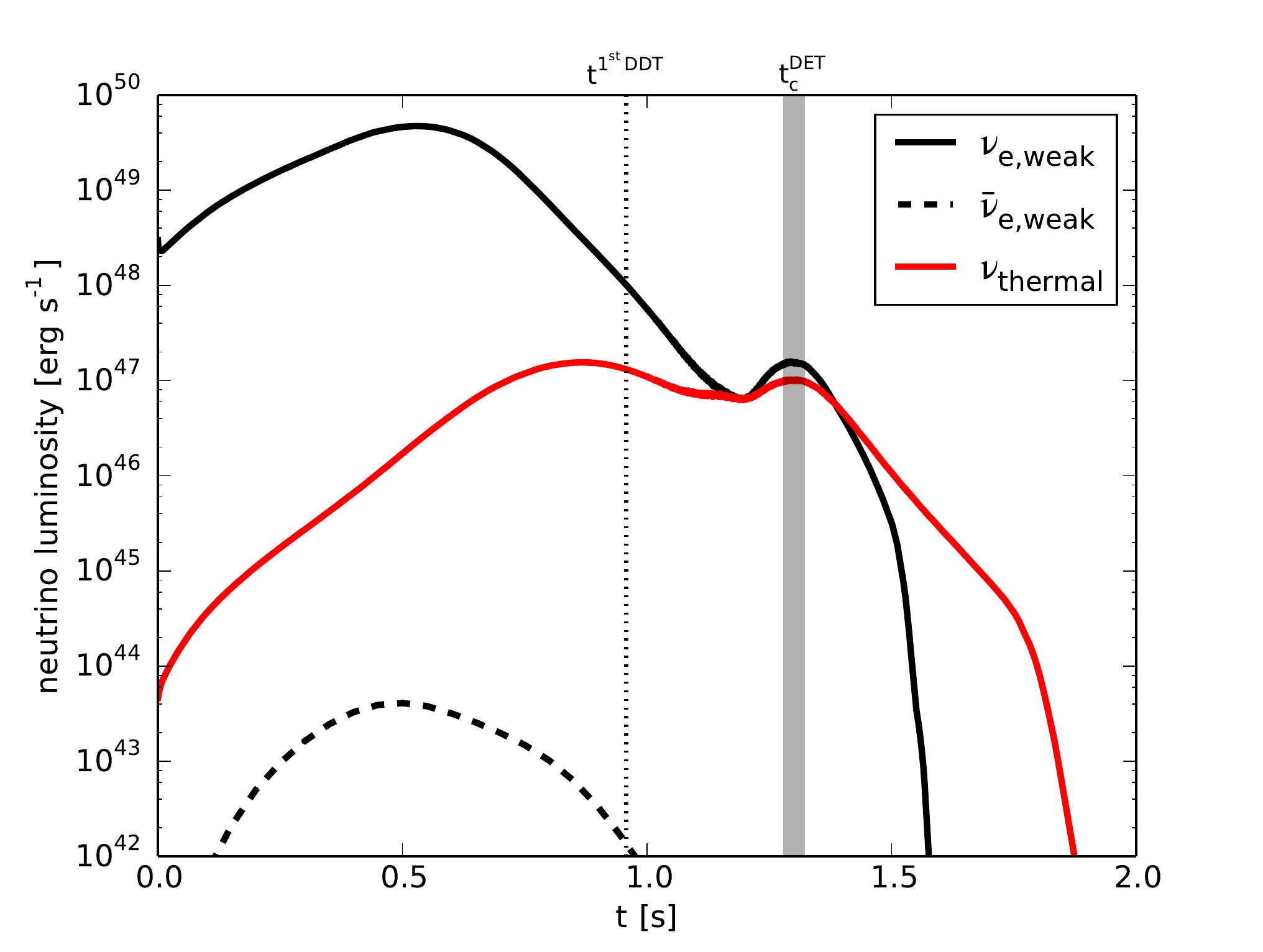}
  \caption{Neutrino luminosity of the N100${\nu}$ model as a function of time,
    originating from weak and thermal neutrinos, respectively. Indicated are the
    time of the first deflagration-to-detonation transition $t^{1^{st}DDT}$ and the
    time $t^{DET}_c$ when the detonation wave reaches the center of the WD.}
  \label{fig:f1}
\end{figure}

\begin{figure}[h]
  \includegraphics[width=\plotsize\columnwidth]{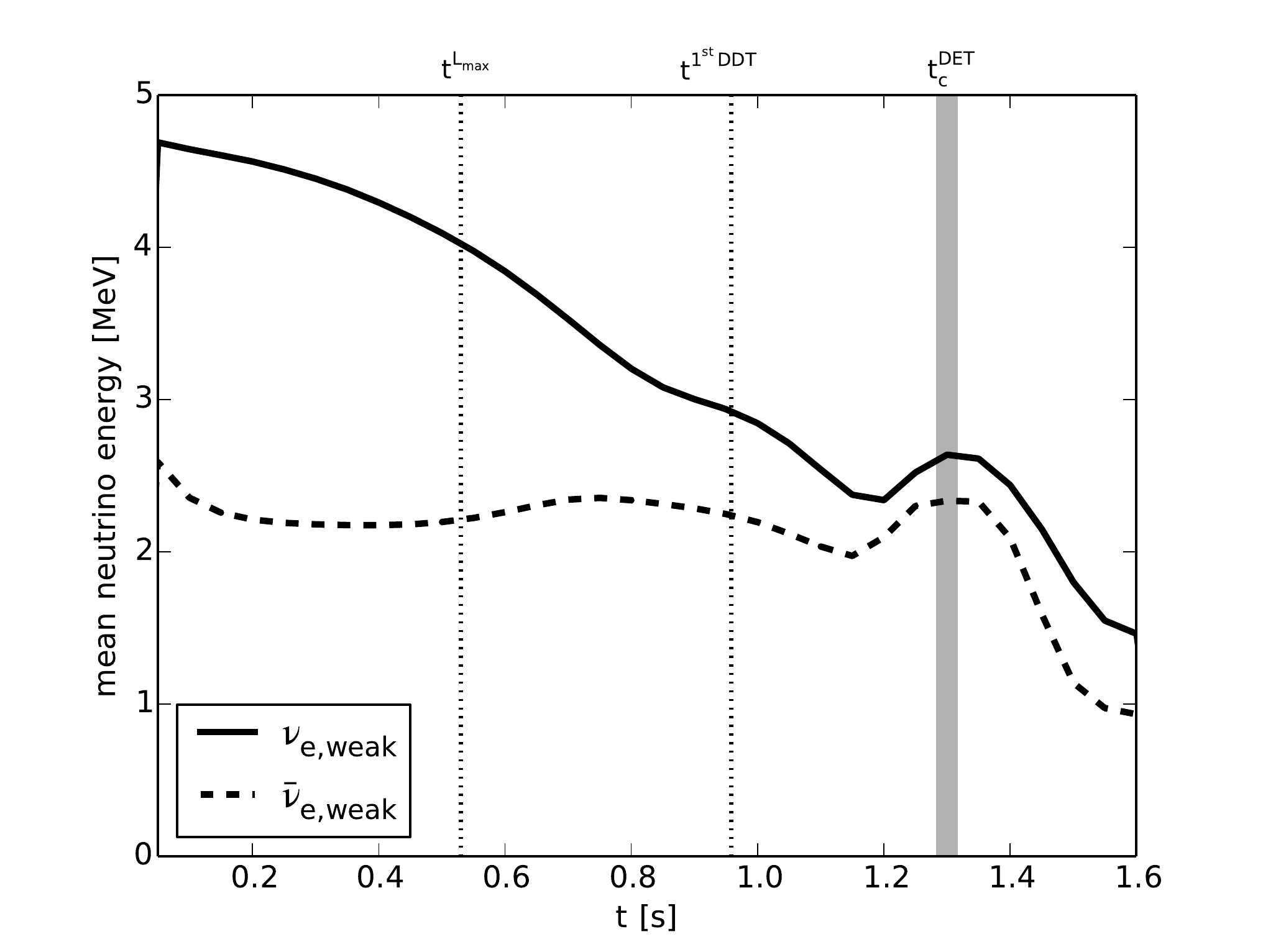}
  \caption{Mean weak neutrino energies of the N100${\nu}$ model as a function of time,
    originating from weak $\nu_e$ and $\bar{\nu}_e$, respectively. Indicated are the
    time of maximum weak neutrino luminosity $t^{L_max}$, the time of the first deflagration-to-detonation transition $t^{1^{st}DDT}$, and the time $t^{DET}_c$ when the detonation wave reaches the center of the WD.}
  \label{fig:f2}
\end{figure}

The weak- and thermal neutrino luminosities for N100${\nu}$
are shown as a function of time in Fig.~\ref{fig:f1}.
The weak neutrino luminosity reaches a maximum of
$4.74\times10^{49}\,\mathrm{erg\,s^{-1}}$ 
at $t=0.53\,\mathrm{s}$ and the integrated total weak neutrino energy loss
amounts to $1.95\times10^{49}\,\mathrm{erg}$. The weak neutrino luminosity is overwhelmingly
dominated by electron neutrinos from electron captures and $\beta^+$ decay. 
In comparison, the electron antineutrino luminosity from positron capture and $\beta^-$ decay is
more than six orders of magnitude smaller (see Fig.~\ref{fig:f1}).
At the time of maximum luminosity, the mean weak electron neutrino energy is 4.0 MeV and the mean weak electron 
antineutrino energy is 2.2 MeV (see Fig.~\ref{fig:f2}).
The thermal neutrino luminosity reaches a maximum of
$1.56\times10^{47}\,\mathrm{erg\,s^{-1}}$ 
at $t=0.87\,\mathrm{s}$ and the integrated total thermal neutrino energy loss
amounts to $8.70\times10^{46}\,\mathrm{erg}$.
The energy drained from the explosion by neutrinos is therefore small compared
to e.g.\ the nuclear energy released in the explosion, the initial gravitational
binding energy, or the final kinetic energy.

Owing to the stochastic nature of our implementation of the
deflagration-to-detonation transition (cf.\ \citep{ciaraldi2013a}), even a small
additional energy loss could, in principle, lead to very different final
explosion. This is, however, not the case here as the time at which the first
DDT occurs is only slightly delayed ($t_{N100}^{1^{st}DDT} =0.928\,\mathrm{s}$
vs. $t_{N100{\nu}}^{1^{st}DDT} =0.958\,\mathrm{s}$).
As a result, there is only a small reduction in kinetic energy between N100
($E_{kin} = 1.45\times10^{51}\,\mathrm{erg}$) and N100${\nu}$
($E_{kin} = 1.42\times10^{51}\,\mathrm{erg}$).
Although dynamically not very important, it is interesting to note that the
energy carried away by neutrinos in ${\sim}1\,\mathrm{s}$ is still on the same
order of magnitude as the total energy radiated in photons by the SN during its
entire duration.

The secondary peak around $1.3\,\mathrm{s}$ in both the weak and thermal
component of the neutrino signal is a signature of the deflagration-to-detonation
transition. The secondary maximum occurs with a time delay with respect to
$t_{N100{\nu}}^{1^{st}DDT}$ since the increased neutrino emission occurs only when
the detonation (which was launched at low density) overruns the highest density
fuel in the center that was left unburned by the deflagration.
\citet{odrzywolek2011a} also find a secondary maximum for the Y12 model and call
it a ``smoking gun'' of delayed-detonation SNe. Their secondary maximum,
however, occurs with a relatively long time delay
($\delta t {\gtrsim}3\,\mathrm{s}$) after the primary maximum. This difference
of the occurrence time of the secondary maximum is a reflection of the
fundamentally different mechanisms that lead to the initiation of the detonation
in the Y12 and N100${\nu}$ models. For a similar reason, these two models also
differ by a factor of ${\sim}6$ in their maximum neutrino luminosities,
$L_v^{max}(\mathrm{Y12}) = 8.2\times10^{48}\,\mathrm{erg\,s^{-1}}$ vs.
$L_v^{max}(\mathrm{N100}\nu) = 4.7\times10^{49}\,\mathrm{erg\,s^{-1}}$. Although
both Y12 and N100${\nu}$ contain both a deflagration and a subsequent
detonation, the fact that N100${\nu}$ burns significantly more mass than Y12
during the (for neutrino emission dominant) deflagration phase leads to the fact
that the N100${\nu}$ neutrino signal is more like that of pure deflagration
models ($L_v^{max}(\mathrm{n7d1r10t15c}) = 3.9\times10^{49}\,\mathrm{erg\,s^{-1}}$
\citep{odrzywolek2011a},
$L_v^{max}(\mathrm{W7}) = 7\times10^{49}\,\mathrm{erg\,s^{-1}}$
\citep{nomoto1984a}), which, however, lack the characteristic
secondary maximum of our simulations.

\section{Gravitational wave signal}
Research on gravitational wave signals of \iae\ is
a largely unexplored field. A large number (${\sim}10^{4}$) Galactic binary
systems -- mostly double white dwarfs -- are expected to be resolved with
future space-based gravitational wave observatories such as {\em eLISA}
\citep{elisa2013a}. Both \citet{nelemans2001a} and \citet{ruiter2010a} predicted
there are about 500 Galactic double degenerate \iae\ progenitors\footnote{These
  double CO white dwarf systems are assumed to be \iae\ progenitors because their
  combined mass exceeds the Chandrasekhar mass limit and they are expected to
  merge within a Hubble time.} that harbour the physical properties (distance,
mass, separation) to enable their detection in gravitational waves.
\citet{dan2011a} calculated the gravitational wave signal of a close
binary system of two WDs in the ringdown phase until the
start of the merger, but they did not model the following thermonuclear
burning which may lead to a supernova. Pioneering work on the signal from an
actual explosion has recently been performed by
\citet{falta2011a} and \citet{falta2011b}, who calculated the
gravitational wave signal of \emph{gravitationally-confined detonation
  (GCD)} models of thermonuclear supernovae (the Y12 model discussed extensively
in Sec.~\ref{sec:nu} falls into this category).
In these models, an off-center ignited deflagration bubble rises
due to buoyancy forces until it
reaches the surface of the WD; the deflagration does not
unbind the star but triggers a subsequent detonation that ignites
opposite to the point where the deflagration bubble reaches the
surface. This scenario leads to asymmetric explosions and thus to
relatively strong gravitational wave signals, which are discussed in
detail in \citep{falta2011a}. 

Based on their earlier results, \citep{falta2011b} analyzed the
stochastic gravitational wave background originating from
thermonuclear supernovae arising from the gravitational confined detonation explosion mechanism. They found that it might pose a considerable
source of noise -- in the frequency range between $0.1\,\mathrm{Hz}$
and $10\,\mathrm{Hz}$ -- for future gravitational wave detectors such
as DECIGO and BBO \citep{kawamura2011a,yagi2011a} that are designed
to detect the gravitational wave signal of cosmological inflation.

\subsection{Numerical implementation}\label{sec:code_gw}
Our numerical approach for the computation of the gravitational wave
signal follows \citet{blanchet1990a} and \citet{mueller1997a}.
We calculate only the approximate \emph{quadrupole radiation} because of the
Newtonian nature of our simulations.

The most straightforward way to determine the amplitude of
gravitational quadrupole waves is by calculating the second derivative
of the quadrupole moment $\mathbf{Q}$.  The gravitational quadrupole
radiation field in the transverse-traceless gauge
$\mathbf{h}^\mathrm{TT}$ is
\begin{align}
  h_{ij}^\mathrm{TT}\left(\mathbf{x},t\right)& =\frac{2G}{c^4D}P_{ijkl}
  \left(\mathbf{n}\right)\frac{\partial^2}
{\partial t^2}Q_{kl}\left(t-\frac{D}{c}\right).\label{eq:htt_quad}
\end{align}
$\mathbf{P}$ is the transverse-traceless projection operator
\begin{align}\label{eq:pijkl} 
  P_{ijkl}\left(\mathbf{n}\right) = &\left(\delta_{ik}-n_in_k\right)
  \left(\delta_{jl}-n_jn_l\right) \\ \nonumber
  & -\frac{1}{2}\left(\delta_{ij}-n_in_j\right)\left(\delta_{kl}-n_kn_l\right),
\end{align}
with the normalized position vector $\mathbf{n}=\mathbf{x}/D$ and the
distance to the source $D=\left|\mathbf{x}\right|$.
However, the calculation of the
second time derivative gives rise to numerical instabilities, the
resulting signal is quite noisy. Therefore, to avoid this unfavorable
method, \citet{nakamura1989a} and \citet{blanchet1990a} introduced a
different way to compute the amplitude. With this new method, which
was applied successfully to core-collapse supernovae by
\citet{moenchmeyer1991a} and \citet{mueller1997a}, the gravitational
quadrupole radiation field $\mathbf{h}^\mathrm{TT}$ is calculated by
\begin{align}
  h_{ij}^\mathrm{TT}\left(\mathbf{x},t\right)& =\frac{2G}{c^4R}P_{ijkl}
  \left(\mathbf{n}\right) \\ \nonumber
  & \int\mathrm{d}^3x\,\rho\left(2v_kv_l-x_k\partial_l
  \Phi- x_l\partial_k\Phi\right).\label{eq:htt}
\end{align}

Here $\mathbf{v}$ 
is the velocity and $\Phi$ is the usual Newtonian
gravitational potential or an effective relativistic gravitational
potential.  \citet{blanchet1990a}
showed that equation~(\ref{eq:htt}) is equivalent to
equation~(\ref{eq:htt_quad}), while \citet{moenchmeyer1991a}
demonstrated the numerical superiority of the method that implements
equation~(\ref{eq:htt}).

Since gravitational waves have two polarization states (``+'' and
``$\times$''), the amplitude can be
written in terms of the two \textit{unit linear-polarization tensors}
$\mathbf{e}_+$ and $\mathbf{e}_\times$ \citep{misner1973a} as
\begin{align}
  h_{ij}^\mathrm{TT}\left(\mathbf{x},t\right)& =\frac{1}{R}
  \left(A_+\mathbf{e}_++A_\times\mathbf{e}_\times\right).
\end{align}
In the case of a plane wave propagating in $z$-direction, the unit
linear-polarization tensors are
\begin{align}
\mathbf{e}_+& =\mathbf{e}_x\otimes\mathbf{e}_x-\mathbf{e}_y\otimes\mathbf{e}_y\\
\mathbf{e}_\times& =\mathbf{e}_x\otimes\mathbf{e}_y+\mathbf{e}_y
\otimes\mathbf{e}_x,
\end{align} 
where $\mathbf{e}_x$ and $\mathbf{e}_y$ are the unit vectors parallel
to the $x$- and $y$-axis, respectively.  We calculate the amplitudes
$A_+$ and $A_\times$ considering two different lines of sight, as was
suggested in \citet{mueller1997a}.  Since they use preferentially
spherical coordinates, they speak of the polar ($\vartheta=0$,
$\varphi=0$) and equatorial ($\vartheta=\pi/2$,~$\varphi=0$)
direction.  In our case, as we use exclusively Cartesian coordinates,
it is more appropriate to call it the $z$-direction and $x$-direction,
which we do from now on.
With the definition 
\begin{align}
  A_{ij}&=\frac{G}{c^4}\int\mathrm{d}^3x\,\rho
  \left(2v_iv_j-x_i\partial_j\Phi-x_j\partial_i\Phi\right),\label{eq:idotdot}
\end{align}
the amplitudes $A_+$ and $A_\times$ in the $z$-direction can be
expressed as
\begin{align}
A_+^z& =A_{xx}-A_{yy}\label{eq:apluspol}\\
A_\times^z& =2A_{xy},
\end{align}
while in the $x$-direction one obtains
\begin{align}
A_+^x&=A_{zz}-A_{yy}\label{eq:apluseq}\\
A^x_\times&=-2A_{yz}\label{eq:akreuzeq}.
\end{align}
We implemented the calculation of these amplitudes according to
equation~(\ref{eq:idotdot}) in the simulation code.
Next we show results of the calculated gravitational wave signal,
beginning with a discussion of the four amplitudes
(\ref{eq:apluspol}--\ref{eq:akreuzeq}).
\subsection{Gravitational wave amplitudes}\label{sec:ia_gwamp}

\begin{figure}
\centering
\includegraphics[width=\plotsize\columnwidth]{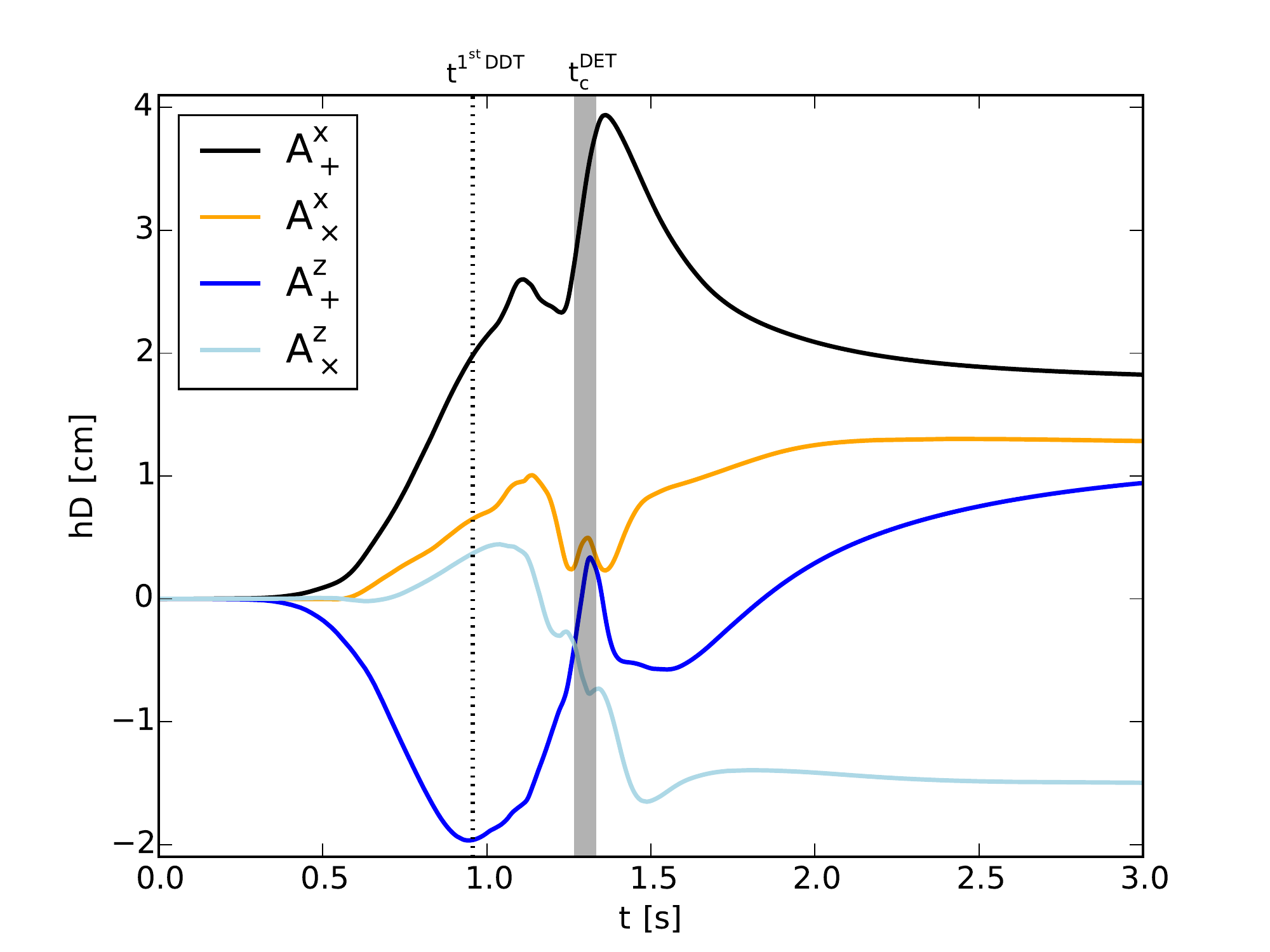}
\caption{Time evolution of four gravitational wave amplitudes for the
  delayed-detonation model N100${\nu}$, see text for details. Indicated are the
  time of the first deflagration-to-detonation transition $t^{1^{st}DDT}$ and the
  time $t^{DET}_c$ when the detonation wave reaches the center of the WD.}
\label{fig:dr14}
\end{figure}

In Fig.~\ref{fig:dr14} we plot four different gravitational wave
amplitudes that arise from two different lines of sight ($x$- and
$z$-direction), each with two polarization states ($\times$ and $+$),
see Sec.~\ref{sec:code_gw} for a detailed explanation.  In the
first ${\sim}0.4\,\mathrm{s}$ after the ignition of the deflagration, no signal
is visible;
after that the absolute values of all amplitudes rise in a steady, monotonic way.
The initiation of the first detonation at $t^{1^{st}DDT}=0.958\,\mathrm{s}$ and the
time $t^{DET}_c{\sim}1.3\,\mathrm{s}$ when the detonation
reaches the center of the WD  leave visible imprints on the signal. For $A_+^x$,
the signal reaches a maximum value of $hD\sim4\,\mathrm{cm}$ at
$t\sim1.36\,\mathrm{s}$, about a factor of $3$ less than the maximum amplitude
obtained by \citep{falta2011a} in their GCD model. The smaller maximum amplitude
of the N100${\nu}$ delayed-detonation model is naturally explained by the
greater compactness of the GCD model when the detonation is triggered.  
The homologous expansion of the supernova ejecta after ${\sim}3\,\mathrm{s}$
generates a signal that is constant in time up to $t=100\,\mathrm{s}$, when the
simulation was stopped.

\subsection{Gravitational wave luminosity and spectrum}
\begin{figure}
\centering
\includegraphics[width=\plotsize\columnwidth]{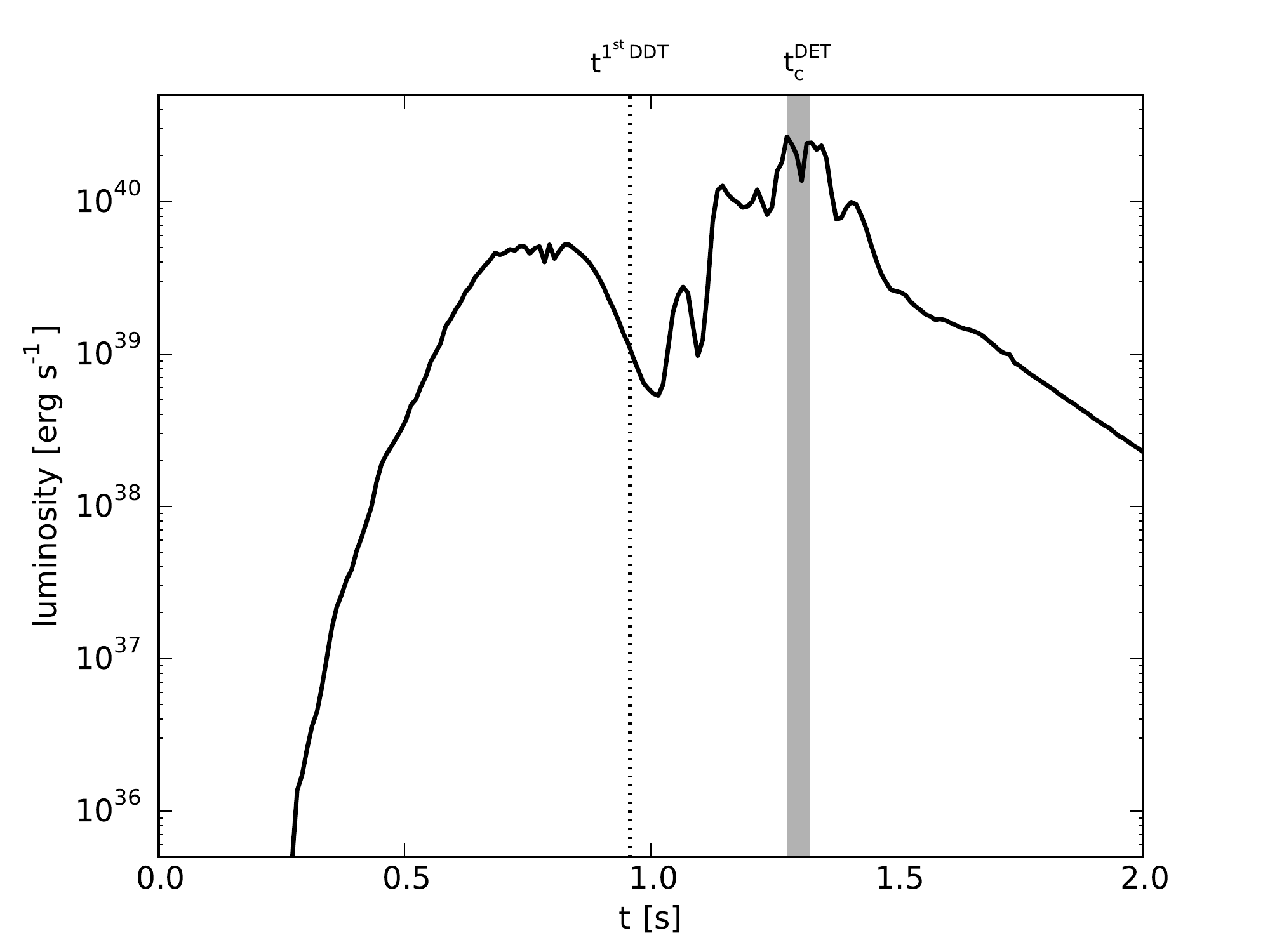}
\caption{Gravitational wave luminosity as a function of time of the
  delayed-detonation model N100${\nu}$. Indicated are the time of the first
  deflagration-to-detonation transition $t^{1^{st}DDT}$ and the time $t^{DET}_c$
  when the detonation wave reaches the center of the WD.}
\label{fig:gwlum}
\end{figure}
\begin{figure}
\centering
\includegraphics[width=\plotsize\columnwidth]{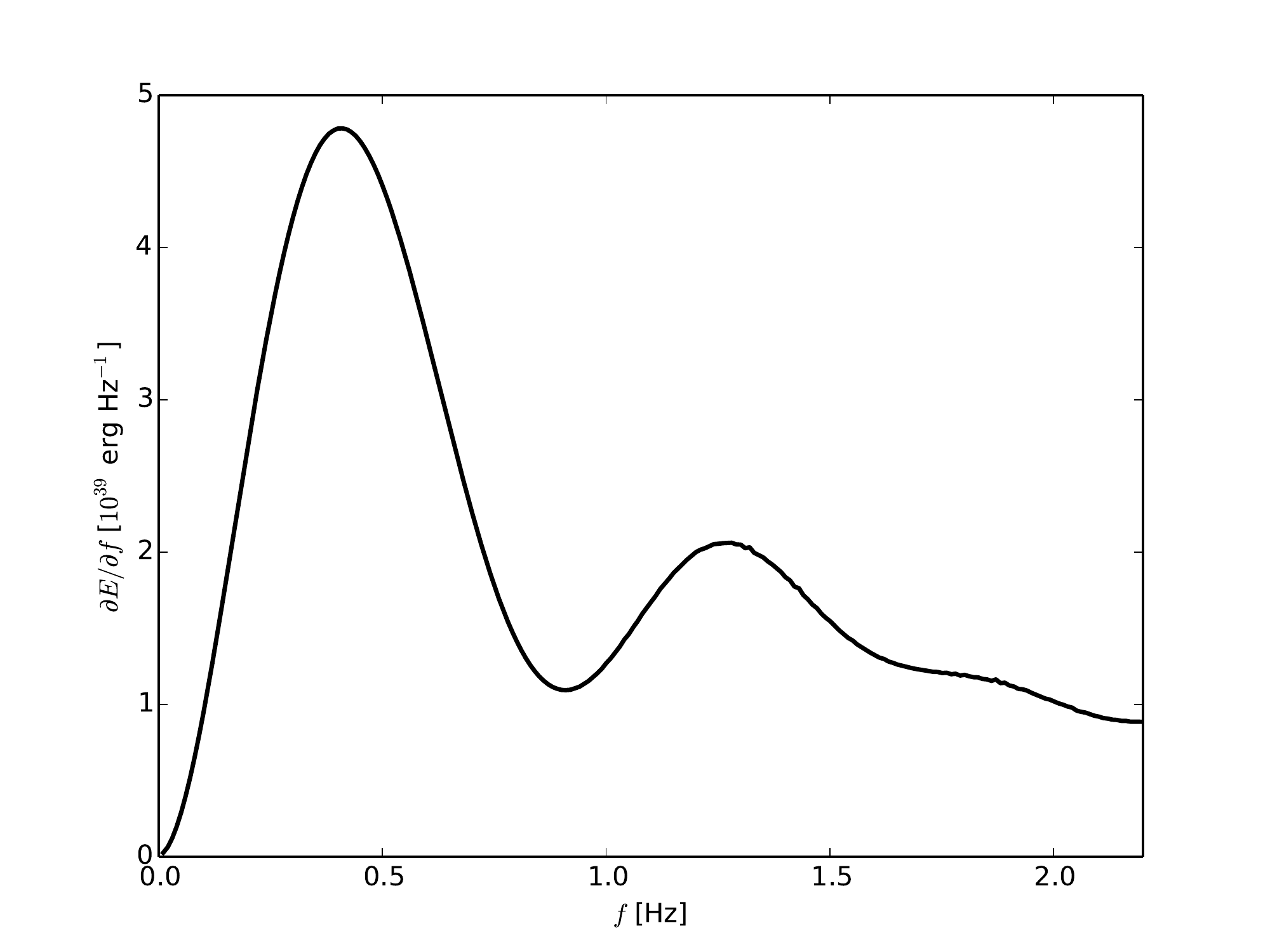}
\caption{Energy spectrum of the gravitational wave signal of the
  delayed-detonation model N100${\nu}$.}
\label{fig:dr23b}
\end{figure}

Beyond the amplitudes, another quantity of interest is the total
amount of energy radiated away by gravitational waves,
$E_\mathrm{gw}$.  It can be expressed as \citep{mueller1997a}
\begin{align}
  \label{eq:egw} \nonumber E_\mathrm{gw}&=\frac{2c^3}{5G}\int^{+\infty}_{-\infty}
  \left(\frac{\mathrm{d}}{\mathrm{d}t}\left(A_{ij}-\frac{1}{3}\delta_{ij}A_{kk}
  \right)\right)^2\mathrm{d}t\\ 
\begin{split}
& = \frac{2c^3}{15G}\int^{+\infty}_{-\infty}\dot
  A^2_{xx}+\dot A^2_{yy}+\dot A^2_{zz}-
  \dot A_{xx}\dot A_{yy}-\dot A_{xx}\dot A_{zz}\\ 
& -\dot A_{yy}\dot A_{zz}
 +3\left(\dot A^2_{xy}+\dot A^2_{xz}+\dot
A^2_{yz}\right)\mathrm{d}t.
\end{split}
\end{align}
All values of $E_\mathrm{gw}$ stated in this work are
calculated by means of equation~(\ref{eq:egw}).

The total gravitational wave energy of N100${\nu}$ amounts to
$6.9\times10^{39}\,\mathrm{erg}$, about an
order of magnitude less than the total energy of the GCD
model of \citet{falta2011a} -- this is again consistent with the results
shown in Sec.~\ref{sec:ia_gwamp}, because the calculated amplitudes
differ by a factor of ${\sim}3$ and the energy is proportional to the square
of the amplitudes. These results demonstrate that the gravitational
wave energy is dynamically unimportant since it is smaller than the internal-,
kinetic-, nuclear-, and even neutrino loss energies. Hence, the error
caused by the fact that we did not couple the gravitational waves part
of the code to the hydrodynamics part is negligible.

The gravitational wave luminosity as a
function of time is plotted in Fig.~\ref{fig:gwlum}. Clearly visible
is a two-peaked structure; the first peak corresponds to the
deflagration phase and features a maximum luminosity of
$8\times10^{39}\,\mathrm{erg\,s^{-1}}$ at $t\sim0.8\,\mathrm{s}$,
while the second, considerably higher peak can be attributed to the
detonation phase, peaking at the time $t^{DET}_c{\sim}1.3\,\mathrm{s}$ when the
detonation reaches the center of the WD and converts the last remaining high
density fuel to iron group elements. Here the maximum luminosity is
$3\times10^{40}\,\mathrm{erg\,s^{-1}}$. In contrast, the GCD model of
\citet{falta2011a} predicts a much greater difference
in maximum power radiated in the deflagration and detonation phases. This
difference is due to the
more weakly ignited deflagration, which leads to a smaller degree of expansion
and correspondingly a detonation in a more compact WD.

The \emph{frequency} $f$ of a gravitational wave signal is an
important quantity to determine, because detectors are only sensitive
in a confined frequency range; see \citep{moore2015a} for a discussion on
quantifying gravitational wave source amplitude and sensitivity curves. We adopt
the approach of \citet{mueller1982b} and \citet{mueller2012b} and conduct Fourier
analyses of the calculated gravitational wave amplitudes and determine
the energy spectrum $\partial E_{gw}/\partial f$. The
Fourier transforms of the $A_{ij}\left(t\right)$ are
\begin{align}
  \label{eq:fourier}
  \tilde{A}_{ij}\left(f\right)&=\int^\infty_{-\infty}A_{ij}
  \left(t\right)\mathrm{e}^{2\pi\mathrm{i}ft}\mathrm{d}t.
\end{align}
Following \citet{mueller2012b}, we obtain the energy spectrum by
\begin{align}
\label{eq:gw_dedf} \nonumber 
\frac{\partial E_{gw}}{\partial f}&=\frac{2c^3}{15G}(2\pi
f)^2 \Bigg(\left|\tilde{A}_{xx}-\tilde{A}_{yy}\right|^2+
\left|\tilde{A}_{xx}-\tilde{A}_{zz}\right|^2 +\\ 
&\left|\tilde{A}_{yy}-\tilde{A}_{zz}\right|^2 
+6\Big(\left|\tilde{A}_{xy}\right|^2 +
\left|\tilde{A}_{xz}\right|^2 + \left|\tilde{A}_{yz}\right|^2\Big)\Bigg)
\end{align}

Fig.~\ref{fig:dr23b} shows the energy spectrum of the gravitational
wave signal, calculated by means of equation~(\ref{eq:gw_dedf}) by evaluating
the Fourier integrals (\ref{eq:fourier}). The spectrum has a pronounced
maximum of $4.8\times10^{39}\,\mathrm{erg}\,\mathrm{s}^{-1}$ at
$f_{max}=0.41\,\mathrm{Hz}$ and a secondary local maximum of
$2.1\times10^{39}\,\mathrm{erg}\,\mathrm{s}^{-1}$ at $1.28\,\mathrm{Hz}$. 

The energy spectrum calculated by
\citep{falta2011a} for the GCD model peaked at ${\sim}2\,\mathrm{Hz}$.
It is worth noting that \citet{falta2011a} stated that an
explosion model with more ``pre-expansion'' prior to detonation, such as
the N100${\nu}$ model presented here, would exhibit a characteristic
gravitational wave frequency shifted downward by a factor of ${\sim}2$,
a prediction that we essentially confirm.

\subsection{Discussion}
\label{sec:discussion}
Our results show that the detection of gravitational wave signals of
\iae{} provides a way to differentiate between different explosion
models with future gravitational wave observatories,
if data analysis techniques are developed to extract those signals.
The signal of our explosion model bears the mark of the
deflagration-to-detonation transition. Gravitational waves generated
by mergers of two WDs would feature the characteristic signature of
the ringdown phase -- according to \citet{dan2011a}, the gravitational
wave signal in the ringdown phase is of the same magnitude as the
signal of the supernova in the above discussed scenarios:
$hD=2\,\mathrm{cm}$ to $20\,\mathrm{cm}$, depending on the WD masses.
Finally, GCD models show a sharp single peak in the time
evolution of the amplitude (figure 2 in \citealt{falta2011a}). However,
the low frequency range of the signals prevents a detection by
second-generation instruments comparable to advanced LIGO; but
third-generation space-based detectors like BBO or
DECIGO that cover a lower frequency range should be able to detect the
gravitational wave signal of at least all Galactic \liae{}.

The gravitational wave frequencies that we obtain are consistent with
the results of \citet{falta2011a} and \citet{falta2011b}. Our results
show that also thermonuclear supernovae modeled as delayed
detonations contribute to the stochastic gravitational wave background
in the frequency range between $0.1\,\mathrm{Hz}$ and
$10\,\mathrm{Hz}$ and thus pose a source of noise that might obscure
the gravitational wave signal originating from the era of
inflation shortly after the Big Bang; the measurement of this signal
is a central goal of the planned detectors BBO and DECIGO.

Finally, we estimate the detectability of our N100$\nu$ delayed-detonation model.
Depending on viewing angle and polarization, our gravitational wave amplitudes
fall in the range between approximately 1 to $4\,\mathrm{cm}$
(see Fig.~\ref{fig:dr14}). At ${\sim}1\,\mathrm{Hz}$, both BBO and DECIGO have a
planned strain sensitivity limit of $h_{min}\approx10^{-24}$ \citep{moore2015a}.
Setting $D_{max} = {hD}/{h_{min}}$ we find that our model's signal would be
detectable out to a distance of
320 kpc and 1.3 Mpc for $A^x_\times$ and $A^x_+$, respectively. Thus, a
delayed-detonation SN~Ia going off in our Milky Way or one of its satellite
galaxies would be well within the sensitivity limits of BBO or DECIGO. For
fortuitous alignment (i.e.\ $A^x_+$), even the Triangulum and Andromeda galaxies
(${\sim}900\,\mathrm{kpc}$ and ${\sim}780\,\mathrm{kpc}$, respectively) would be
within reach.

\section{Conclusions}\label{sec:conclusions}
We conducted a three-dimensional hydrodynamic simulation of a
delayed-detonation explosion of a thermonuclear supernova and computed the
neutrino and gravitational wave signals. By taking the energetics of neutrino
losses on the hydrodynamic evolution into account, we find a reduction in kinetic
energy of about two per cent and we conclude that the effects of neutrino
emission on the explosion dynamics are indeed small. Nevertheless, our model
emits $2\times10^{49}\,\mathrm{erg}$ in neutrinos, comparable to the total energy
radiated by photons. We present the amplitudes, luminosity as a function of time,
and the energy spectrum of the gravitational wave signal of our SN~Ia supernova
explosion simulation. The energy spectrum of the gravitational wave emission
peaks where the proposed space-based gravitational wave missions \mbox{DECIGO}
and BBO have the highest design sensitivity and we find that the signal is
potentially observable out to a distance of 1.3\,Mpc. In both the neutrino and
gravitational wave signals we find a distinct signature of the
deflagration-to-detonation transition, which, in principle, makes it
possible to constrain the explosion model of SNe~Ia by observations of their
non-electromagnetic neutrino and/gravitational wave radiation.

\acknowledgments{
  We thank Chistopher Moore for his insightful comments on sensitivity curves of
  gravitational wave detectors and R\"udiger Pakmor for
  their discussions on power spectra and Fourier transforms. IRS was funded by
  the Australian Research Council Laureate Grant FL0992131 and the Deutsche
  Forschungsgemein\-schaft (DFG) through the graduate school on ``Theoretical
  Astrophysics and Particle Physics'' (GRK 1147), MH by CompStar, a Research
  Networking Programme of the European Science Foundation, and FKR by the DFG
  via the Emmy Noether Program (RO 3676/1-1) and by the ARCHES prize of the
  German Federal Ministry of Education and Research (BMBF). AJR acknowledges
  funding support from the Australian Research Council Centre of Excellence
  for All-sky Astrophysics (CAASTRO) through project number CE110001020.
  STO acknowledges support from the Studienstiftung des Deutschen Volkes. 
  Funding for collaboration was provided by the DAAD/Go8 German-Australian
  exchange program. This research was supported by the Partner Time Allocation
 (Australian National University) and the National Computational Merit Allocation
 Schemes of the NCI National Facility at the Australian National University. 
 Part of the simulations were carried out on the JUGENE supercomputer at the 
 Forschungszentrum J{\"u}lich within the Partnership for Advanced Computing in Europe (PRA042).
}


\begin{thebibliography}{74}
\expandafter\ifx\csname natexlab\endcsname\relax\def\natexlab#1{#1}\fi
\expandafter\ifx\csname bibnamefont\endcsname\relax
  \def\bibnamefont#1{#1}\fi
\expandafter\ifx\csname bibfnamefont\endcsname\relax
  \def\bibfnamefont#1{#1}\fi
\expandafter\ifx\csname citenamefont\endcsname\relax
  \def\citenamefont#1{#1}\fi
\expandafter\ifx\csname url\endcsname\relax
  \def\url#1{\texttt{#1}}\fi
\expandafter\ifx\csname urlprefix\endcsname\relax\def\urlprefix{URL }\fi
\providecommand{\bibinfo}[2]{#2}
\providecommand{\eprint}[2][]{\url{#2}}

\bibitem[{\citenamefont{{Kasen} et~al.}(2009)\citenamefont{{Kasen},
  {R{\"o}pke}, and {Woosley}}}]{kasen2009a}
\bibinfo{author}{\bibfnamefont{D.}~\bibnamefont{{Kasen}}},
  \bibinfo{author}{\bibfnamefont{F.~K.} \bibnamefont{{R{\"o}pke}}},
  \bibnamefont{and} \bibinfo{author}{\bibfnamefont{S.~E.}
  \bibnamefont{{Woosley}}}, \bibinfo{journal}{\nat}
  \textbf{\bibinfo{volume}{460}}, \bibinfo{pages}{869} (\bibinfo{year}{2009}),
  \eprint{0907.0708}.

\bibitem[{\citenamefont{{Kromer} et~al.}(2010)\citenamefont{{Kromer}, {Sim},
  {Fink}, {R{\"o}pke}, {Seitenzahl}, and {Hillebrandt}}}]{kromer2010a}
\bibinfo{author}{\bibfnamefont{M.}~\bibnamefont{{Kromer}}},
  \bibinfo{author}{\bibfnamefont{S.~A.} \bibnamefont{{Sim}}},
  \bibinfo{author}{\bibfnamefont{M.}~\bibnamefont{{Fink}}},
  \bibinfo{author}{\bibfnamefont{F.~K.} \bibnamefont{{R{\"o}pke}}},
  \bibinfo{author}{\bibfnamefont{I.~R.} \bibnamefont{{Seitenzahl}}},
  \bibnamefont{and}
  \bibinfo{author}{\bibfnamefont{W.}~\bibnamefont{{Hillebrandt}}},
  \bibinfo{journal}{\apj} \textbf{\bibinfo{volume}{719}}, \bibinfo{pages}{1067}
  (\bibinfo{year}{2010}), \eprint{1006.4489}.

\bibitem[{\citenamefont{{Pakmor} et~al.}(2012)\citenamefont{{Pakmor}, {Kromer},
  {Taubenberger}, {Sim}, {R{\"o}pke}, and {Hillebrandt}}}]{pakmor2012a}
\bibinfo{author}{\bibfnamefont{R.}~\bibnamefont{{Pakmor}}},
  \bibinfo{author}{\bibfnamefont{M.}~\bibnamefont{{Kromer}}},
  \bibinfo{author}{\bibfnamefont{S.}~\bibnamefont{{Taubenberger}}},
  \bibinfo{author}{\bibfnamefont{S.~A.} \bibnamefont{{Sim}}},
  \bibinfo{author}{\bibfnamefont{F.~K.} \bibnamefont{{R{\"o}pke}}},
  \bibnamefont{and}
  \bibinfo{author}{\bibfnamefont{W.}~\bibnamefont{{Hillebrandt}}},
  \bibinfo{journal}{\apjl} \textbf{\bibinfo{volume}{747}}, \bibinfo{eid}{L10}
  (\bibinfo{year}{2012}), \eprint{1201.5123}.

\bibitem[{\citenamefont{{Kromer} et~al.}(2013)\citenamefont{{Kromer}, {Fink},
  {Stanishev}, {Taubenberger}, {Ciaraldi-Schoolman}, {Pakmor}, {R{\"o}pke},
  {Ruiter}, {Seitenzahl}, {Sim} et~al.}}]{kromer2013a}
\bibinfo{author}{\bibfnamefont{M.}~\bibnamefont{{Kromer}}},
  \bibinfo{author}{\bibfnamefont{M.}~\bibnamefont{{Fink}}},
  \bibinfo{author}{\bibfnamefont{V.}~\bibnamefont{{Stanishev}}},
  \bibinfo{author}{\bibfnamefont{S.}~\bibnamefont{{Taubenberger}}},
  \bibinfo{author}{\bibfnamefont{F.}~\bibnamefont{{Ciaraldi-Schoolman}}},
  \bibinfo{author}{\bibfnamefont{R.}~\bibnamefont{{Pakmor}}},
  \bibinfo{author}{\bibfnamefont{F.~K.} \bibnamefont{{R{\"o}pke}}},
  \bibinfo{author}{\bibfnamefont{A.~J.} \bibnamefont{{Ruiter}}},
  \bibinfo{author}{\bibfnamefont{I.~R.} \bibnamefont{{Seitenzahl}}},
  \bibinfo{author}{\bibfnamefont{S.~A.} \bibnamefont{{Sim}}},
  \bibnamefont{et~al.}, \bibinfo{journal}{\mnras}
  \textbf{\bibinfo{volume}{429}}, \bibinfo{pages}{2287} (\bibinfo{year}{2013}),
  \eprint{1210.5243}.

\bibitem[{\citenamefont{{Sim} et~al.}(2013)\citenamefont{{Sim}, {Seitenzahl},
  {Kromer}, {Ciaraldi-Schoolmann}, {R{\"o}pke}, {Fink}, {Hillebrandt},
  {Pakmor}, {Ruiter}, and {Taubenberger}}}]{sim2013a}
\bibinfo{author}{\bibfnamefont{S.~A.} \bibnamefont{{Sim}}},
  \bibinfo{author}{\bibfnamefont{I.~R.} \bibnamefont{{Seitenzahl}}},
  \bibinfo{author}{\bibfnamefont{M.}~\bibnamefont{{Kromer}}},
  \bibinfo{author}{\bibfnamefont{F.}~\bibnamefont{{Ciaraldi-Schoolmann}}},
  \bibinfo{author}{\bibfnamefont{F.~K.} \bibnamefont{{R{\"o}pke}}},
  \bibinfo{author}{\bibfnamefont{M.}~\bibnamefont{{Fink}}},
  \bibinfo{author}{\bibfnamefont{W.}~\bibnamefont{{Hillebrandt}}},
  \bibinfo{author}{\bibfnamefont{R.}~\bibnamefont{{Pakmor}}},
  \bibinfo{author}{\bibfnamefont{A.~J.} \bibnamefont{{Ruiter}}},
  \bibnamefont{and}
  \bibinfo{author}{\bibfnamefont{S.}~\bibnamefont{{Taubenberger}}},
  \bibinfo{journal}{\mnras} \textbf{\bibinfo{volume}{436}},
  \bibinfo{pages}{333} (\bibinfo{year}{2013}), \eprint{1308.4833}.

\bibitem[{\citenamefont{{R{\"o}pke} et~al.}(2012)\citenamefont{{R{\"o}pke},
  {Kromer}, {Seitenzahl}, {Pakmor}, {Sim}, {Taubenberger},
  {Ciaraldi-Schoolmann}, {Hillebrandt}, {Aldering}, {Antilogus}
  et~al.}}]{roepke2012a}
\bibinfo{author}{\bibfnamefont{F.~K.} \bibnamefont{{R{\"o}pke}}},
  \bibinfo{author}{\bibfnamefont{M.}~\bibnamefont{{Kromer}}},
  \bibinfo{author}{\bibfnamefont{I.~R.} \bibnamefont{{Seitenzahl}}},
  \bibinfo{author}{\bibfnamefont{R.}~\bibnamefont{{Pakmor}}},
  \bibinfo{author}{\bibfnamefont{S.~A.} \bibnamefont{{Sim}}},
  \bibinfo{author}{\bibfnamefont{S.}~\bibnamefont{{Taubenberger}}},
  \bibinfo{author}{\bibfnamefont{F.}~\bibnamefont{{Ciaraldi-Schoolmann}}},
  \bibinfo{author}{\bibfnamefont{W.}~\bibnamefont{{Hillebrandt}}},
  \bibinfo{author}{\bibfnamefont{G.}~\bibnamefont{{Aldering}}},
  \bibinfo{author}{\bibfnamefont{P.}~\bibnamefont{{Antilogus}}},
  \bibnamefont{et~al.}, \bibinfo{journal}{\apjl}
  \textbf{\bibinfo{volume}{750}}, \bibinfo{eid}{L19} (\bibinfo{year}{2012}),
  \eprint{1203.4839}.

\bibitem[{\citenamefont{{Stehle} et~al.}(2005)\citenamefont{{Stehle},
  {Mazzali}, {Benetti}, and {Hillebrandt}}}]{stehle2005a}
\bibinfo{author}{\bibfnamefont{M.}~\bibnamefont{{Stehle}}},
  \bibinfo{author}{\bibfnamefont{P.~A.} \bibnamefont{{Mazzali}}},
  \bibinfo{author}{\bibfnamefont{S.}~\bibnamefont{{Benetti}}},
  \bibnamefont{and}
  \bibinfo{author}{\bibfnamefont{W.}~\bibnamefont{{Hillebrandt}}},
  \bibinfo{journal}{\mnras} \textbf{\bibinfo{volume}{360}},
  \bibinfo{pages}{1231} (\bibinfo{year}{2005}),
  \eprint{arXiv:astro-ph/0409342}.

\bibitem[{\citenamefont{{Mazzali} et~al.}(2008)\citenamefont{{Mazzali},
  {Nomoto}, {Maeda}, {Deng}, {Benetti}, {R{\"o}pke}, and
  {Hillebrandt}}}]{mazzali2008a}
\bibinfo{author}{\bibfnamefont{P.~A.} \bibnamefont{{Mazzali}}},
  \bibinfo{author}{\bibfnamefont{K.}~\bibnamefont{{Nomoto}}},
  \bibinfo{author}{\bibfnamefont{K.}~\bibnamefont{{Maeda}}},
  \bibinfo{author}{\bibfnamefont{J.}~\bibnamefont{{Deng}}},
  \bibinfo{author}{\bibfnamefont{S.}~\bibnamefont{{Benetti}}},
  \bibinfo{author}{\bibfnamefont{F.}~\bibnamefont{{R{\"o}pke}}},
  \bibnamefont{and}
  \bibinfo{author}{\bibfnamefont{W.}~\bibnamefont{{Hillebrandt}}},
  \textbf{\bibinfo{volume}{391}}, \bibinfo{pages}{347} (\bibinfo{year}{2008}).

\bibitem[{\citenamefont{{Hachinger} et~al.}(2009)\citenamefont{{Hachinger},
  {Mazzali}, {Taubenberger}, {Pakmor}, and {Hillebrandt}}}]{hachinger2009a}
\bibinfo{author}{\bibfnamefont{S.}~\bibnamefont{{Hachinger}}},
  \bibinfo{author}{\bibfnamefont{P.~A.} \bibnamefont{{Mazzali}}},
  \bibinfo{author}{\bibfnamefont{S.}~\bibnamefont{{Taubenberger}}},
  \bibinfo{author}{\bibfnamefont{R.}~\bibnamefont{{Pakmor}}}, \bibnamefont{and}
  \bibinfo{author}{\bibfnamefont{W.}~\bibnamefont{{Hillebrandt}}},
  \bibinfo{journal}{\mnras} \textbf{\bibinfo{volume}{399}},
  \bibinfo{pages}{1238} (\bibinfo{year}{2009}), \eprint{0907.2542}.

\bibitem[{\citenamefont{{Hachinger} et~al.}(2013)\citenamefont{{Hachinger},
  {Mazzali}, {Sullivan}, {Ellis}, {Maguire}, {Gal-Yam}, {Howell}, {Nugent},
  {Baron}, {Cooke} et~al.}}]{hachinger2013a}
\bibinfo{author}{\bibfnamefont{S.}~\bibnamefont{{Hachinger}}},
  \bibinfo{author}{\bibfnamefont{P.~A.} \bibnamefont{{Mazzali}}},
  \bibinfo{author}{\bibfnamefont{M.}~\bibnamefont{{Sullivan}}},
  \bibinfo{author}{\bibfnamefont{R.~S.} \bibnamefont{{Ellis}}},
  \bibinfo{author}{\bibfnamefont{K.}~\bibnamefont{{Maguire}}},
  \bibinfo{author}{\bibfnamefont{A.}~\bibnamefont{{Gal-Yam}}},
  \bibinfo{author}{\bibfnamefont{D.~A.} \bibnamefont{{Howell}}},
  \bibinfo{author}{\bibfnamefont{P.~E.} \bibnamefont{{Nugent}}},
  \bibinfo{author}{\bibfnamefont{E.}~\bibnamefont{{Baron}}},
  \bibinfo{author}{\bibfnamefont{J.}~\bibnamefont{{Cooke}}},
  \bibnamefont{et~al.}, \bibinfo{journal}{\mnras}
  \textbf{\bibinfo{volume}{429}}, \bibinfo{pages}{2228} (\bibinfo{year}{2013}),
  \eprint{1208.1267}.

\bibitem[{\citenamefont{{Kasen}}(2010)}]{kasen2010a}
\bibinfo{author}{\bibfnamefont{D.}~\bibnamefont{{Kasen}}},
  \bibinfo{journal}{\apj} \textbf{\bibinfo{volume}{708}}, \bibinfo{pages}{1025}
  (\bibinfo{year}{2010}), \eprint{0909.0275}.

\bibitem[{\citenamefont{{Bloom} et~al.}(2012)\citenamefont{{Bloom}, {Kasen},
  {Shen}, {Nugent}, {Butler}, {Graham}, {Howell}, {Kolb}, {Holmes}, {Haswell}
  et~al.}}]{bloom2012a}
\bibinfo{author}{\bibfnamefont{J.~S.} \bibnamefont{{Bloom}}},
  \bibinfo{author}{\bibfnamefont{D.}~\bibnamefont{{Kasen}}},
  \bibinfo{author}{\bibfnamefont{K.~J.} \bibnamefont{{Shen}}},
  \bibinfo{author}{\bibfnamefont{P.~E.} \bibnamefont{{Nugent}}},
  \bibinfo{author}{\bibfnamefont{N.~R.} \bibnamefont{{Butler}}},
  \bibinfo{author}{\bibfnamefont{M.~L.} \bibnamefont{{Graham}}},
  \bibinfo{author}{\bibfnamefont{D.~A.} \bibnamefont{{Howell}}},
  \bibinfo{author}{\bibfnamefont{U.}~\bibnamefont{{Kolb}}},
  \bibinfo{author}{\bibfnamefont{S.}~\bibnamefont{{Holmes}}},
  \bibinfo{author}{\bibfnamefont{C.~A.} \bibnamefont{{Haswell}}},
  \bibnamefont{et~al.}, \bibinfo{journal}{\apjl}
  \textbf{\bibinfo{volume}{744}}, \bibinfo{eid}{L17} (\bibinfo{year}{2012}),
  \eprint{1111.0966}.

\bibitem[{\citenamefont{{Brown} et~al.}(2012)\citenamefont{{Brown}, {Dawson},
  {de Pasquale}, {Gronwall}, {Holland}, {Immler}, {Kuin}, {Mazzali}, {Milne},
  {Oates} et~al.}}]{brown2012a}
\bibinfo{author}{\bibfnamefont{P.~J.} \bibnamefont{{Brown}}},
  \bibinfo{author}{\bibfnamefont{K.~S.} \bibnamefont{{Dawson}}},
  \bibinfo{author}{\bibfnamefont{M.}~\bibnamefont{{de Pasquale}}},
  \bibinfo{author}{\bibfnamefont{C.}~\bibnamefont{{Gronwall}}},
  \bibinfo{author}{\bibfnamefont{S.}~\bibnamefont{{Holland}}},
  \bibinfo{author}{\bibfnamefont{S.}~\bibnamefont{{Immler}}},
  \bibinfo{author}{\bibfnamefont{P.}~\bibnamefont{{Kuin}}},
  \bibinfo{author}{\bibfnamefont{P.}~\bibnamefont{{Mazzali}}},
  \bibinfo{author}{\bibfnamefont{P.}~\bibnamefont{{Milne}}},
  \bibinfo{author}{\bibfnamefont{S.}~\bibnamefont{{Oates}}},
  \bibnamefont{et~al.}, \bibinfo{journal}{\apj} \textbf{\bibinfo{volume}{753}},
  \bibinfo{eid}{22} (\bibinfo{year}{2012}), \eprint{1110.2538}.

\bibitem[{\citenamefont{{Patat} et~al.}(2007)\citenamefont{{Patat}, {Chandra},
  {Chevalier}, {Justham}, {Podsiadlowski}, {Wolf}, {Gal-Yam}, {Pasquini},
  {Crawford}, {Mazzali} et~al.}}]{patat2007a}
\bibinfo{author}{\bibfnamefont{F.}~\bibnamefont{{Patat}}},
  \bibinfo{author}{\bibfnamefont{P.}~\bibnamefont{{Chandra}}},
  \bibinfo{author}{\bibfnamefont{R.}~\bibnamefont{{Chevalier}}},
  \bibinfo{author}{\bibfnamefont{S.}~\bibnamefont{{Justham}}},
  \bibinfo{author}{\bibfnamefont{P.}~\bibnamefont{{Podsiadlowski}}},
  \bibinfo{author}{\bibfnamefont{C.}~\bibnamefont{{Wolf}}},
  \bibinfo{author}{\bibfnamefont{A.}~\bibnamefont{{Gal-Yam}}},
  \bibinfo{author}{\bibfnamefont{L.}~\bibnamefont{{Pasquini}}},
  \bibinfo{author}{\bibfnamefont{I.~A.} \bibnamefont{{Crawford}}},
  \bibinfo{author}{\bibfnamefont{P.~A.} \bibnamefont{{Mazzali}}},
  \bibnamefont{et~al.}, \bibinfo{journal}{Science}
  \textbf{\bibinfo{volume}{317}}, \bibinfo{pages}{924} (\bibinfo{year}{2007}),
  \eprint{arXiv:0707.2793}.

\bibitem[{\citenamefont{{Sternberg} et~al.}(2011)\citenamefont{{Sternberg},
  {Gal-Yam}, {Simon}, {Leonard}, {Quimby}, {Phillips}, {Morrell}, {Thompson},
  {Ivans}, {Marshall} et~al.}}]{sternberg2011a}
\bibinfo{author}{\bibfnamefont{A.}~\bibnamefont{{Sternberg}}},
  \bibinfo{author}{\bibfnamefont{A.}~\bibnamefont{{Gal-Yam}}},
  \bibinfo{author}{\bibfnamefont{J.~D.} \bibnamefont{{Simon}}},
  \bibinfo{author}{\bibfnamefont{D.~C.} \bibnamefont{{Leonard}}},
  \bibinfo{author}{\bibfnamefont{R.~M.} \bibnamefont{{Quimby}}},
  \bibinfo{author}{\bibfnamefont{M.~M.} \bibnamefont{{Phillips}}},
  \bibinfo{author}{\bibfnamefont{N.}~\bibnamefont{{Morrell}}},
  \bibinfo{author}{\bibfnamefont{I.~B.} \bibnamefont{{Thompson}}},
  \bibinfo{author}{\bibfnamefont{I.}~\bibnamefont{{Ivans}}},
  \bibinfo{author}{\bibfnamefont{J.~L.} \bibnamefont{{Marshall}}},
  \bibnamefont{et~al.}, \bibinfo{journal}{Science}
  \textbf{\bibinfo{volume}{333}}, \bibinfo{pages}{856} (\bibinfo{year}{2011}),
  \eprint{1108.3664}.

\bibitem[{\citenamefont{{Dilday} et~al.}(2012)\citenamefont{{Dilday}, {Howell},
  {Cenko}, {Silverman}, {Nugent}, {Sullivan}, {Ben-Ami}, {Bildsten}, {Bolte},
  {Endl} et~al.}}]{dilday2012a}
\bibinfo{author}{\bibfnamefont{B.}~\bibnamefont{{Dilday}}},
  \bibinfo{author}{\bibfnamefont{D.~A.} \bibnamefont{{Howell}}},
  \bibinfo{author}{\bibfnamefont{S.~B.} \bibnamefont{{Cenko}}},
  \bibinfo{author}{\bibfnamefont{J.~M.} \bibnamefont{{Silverman}}},
  \bibinfo{author}{\bibfnamefont{P.~E.} \bibnamefont{{Nugent}}},
  \bibinfo{author}{\bibfnamefont{M.}~\bibnamefont{{Sullivan}}},
  \bibinfo{author}{\bibfnamefont{S.}~\bibnamefont{{Ben-Ami}}},
  \bibinfo{author}{\bibfnamefont{L.}~\bibnamefont{{Bildsten}}},
  \bibinfo{author}{\bibfnamefont{M.}~\bibnamefont{{Bolte}}},
  \bibinfo{author}{\bibfnamefont{M.}~\bibnamefont{{Endl}}},
  \bibnamefont{et~al.}, \bibinfo{journal}{Science}
  \textbf{\bibinfo{volume}{337}}, \bibinfo{pages}{942} (\bibinfo{year}{2012}),
  \eprint{1207.1306}.

\bibitem[{\citenamefont{{Seitenzahl}
  et~al.}(2009{\natexlab{a}})\citenamefont{{Seitenzahl}, {Taubenberger}, and
  {Sim}}}]{seitenzahl2009d}
\bibinfo{author}{\bibfnamefont{I.~R.} \bibnamefont{{Seitenzahl}}},
  \bibinfo{author}{\bibfnamefont{S.}~\bibnamefont{{Taubenberger}}},
  \bibnamefont{and} \bibinfo{author}{\bibfnamefont{S.~A.} \bibnamefont{{Sim}}},
  \bibinfo{journal}{\mnras} \textbf{\bibinfo{volume}{400}},
  \bibinfo{pages}{531} (\bibinfo{year}{2009}{\natexlab{a}}),
  \eprint{0908.0247}.

\bibitem[{\citenamefont{{Kerzendorf} et~al.}(2014)\citenamefont{{Kerzendorf},
  {Taubenberger}, {Seitenzahl}, and {Ruiter}}}]{kerzendorf2014a}
\bibinfo{author}{\bibfnamefont{W.~E.} \bibnamefont{{Kerzendorf}}},
  \bibinfo{author}{\bibfnamefont{S.}~\bibnamefont{{Taubenberger}}},
  \bibinfo{author}{\bibfnamefont{I.~R.} \bibnamefont{{Seitenzahl}}},
  \bibnamefont{and} \bibinfo{author}{\bibfnamefont{A.~J.}
  \bibnamefont{{Ruiter}}}, \bibinfo{journal}{\apjl}
  \textbf{\bibinfo{volume}{796}}, \bibinfo{eid}{L26} (\bibinfo{year}{2014}),
  \eprint{1406.6050}.

\bibitem[{\citenamefont{{Sim} and {Mazzali}}(2008)}]{sim2008a}
\bibinfo{author}{\bibfnamefont{S.~A.} \bibnamefont{{Sim}}} \bibnamefont{and}
  \bibinfo{author}{\bibfnamefont{P.~A.} \bibnamefont{{Mazzali}}},
  \bibinfo{journal}{\mnras} \textbf{\bibinfo{volume}{385}},
  \bibinfo{pages}{1681} (\bibinfo{year}{2008}), \eprint{0710.3313}.

\bibitem[{\citenamefont{{Maeda} et~al.}(2012)\citenamefont{{Maeda}, {Terada},
  {Kasen}, {R{\"o}pke}, {Bamba}, {Diehl}, {Nomoto}, {Kromer}, {Seitenzahl},
  {Yamaguchi} et~al.}}]{maeda2012b}
\bibinfo{author}{\bibfnamefont{K.}~\bibnamefont{{Maeda}}},
  \bibinfo{author}{\bibfnamefont{Y.}~\bibnamefont{{Terada}}},
  \bibinfo{author}{\bibfnamefont{D.}~\bibnamefont{{Kasen}}},
  \bibinfo{author}{\bibfnamefont{F.~K.} \bibnamefont{{R{\"o}pke}}},
  \bibinfo{author}{\bibfnamefont{A.}~\bibnamefont{{Bamba}}},
  \bibinfo{author}{\bibfnamefont{R.}~\bibnamefont{{Diehl}}},
  \bibinfo{author}{\bibfnamefont{K.}~\bibnamefont{{Nomoto}}},
  \bibinfo{author}{\bibfnamefont{M.}~\bibnamefont{{Kromer}}},
  \bibinfo{author}{\bibfnamefont{I.~R.} \bibnamefont{{Seitenzahl}}},
  \bibinfo{author}{\bibfnamefont{H.}~\bibnamefont{{Yamaguchi}}},
  \bibnamefont{et~al.}, \bibinfo{journal}{\apj} \textbf{\bibinfo{volume}{760}},
  \bibinfo{eid}{54} (\bibinfo{year}{2012}), \eprint{1208.2094}.

\bibitem[{\citenamefont{{Summa} et~al.}(2013)\citenamefont{{Summa}, {Ulyanov},
  {Kromer}, {Boyer}, {R{\"o}pke}, {Sim}, {Seitenzahl}, {Fink}, {Mannheim},
  {Pakmor} et~al.}}]{summa2013a}
\bibinfo{author}{\bibfnamefont{A.}~\bibnamefont{{Summa}}},
  \bibinfo{author}{\bibfnamefont{A.}~\bibnamefont{{Ulyanov}}},
  \bibinfo{author}{\bibfnamefont{M.}~\bibnamefont{{Kromer}}},
  \bibinfo{author}{\bibfnamefont{S.}~\bibnamefont{{Boyer}}},
  \bibinfo{author}{\bibfnamefont{F.~K.} \bibnamefont{{R{\"o}pke}}},
  \bibinfo{author}{\bibfnamefont{S.~A.} \bibnamefont{{Sim}}},
  \bibinfo{author}{\bibfnamefont{I.~R.} \bibnamefont{{Seitenzahl}}},
  \bibinfo{author}{\bibfnamefont{M.}~\bibnamefont{{Fink}}},
  \bibinfo{author}{\bibfnamefont{K.}~\bibnamefont{{Mannheim}}},
  \bibinfo{author}{\bibfnamefont{R.}~\bibnamefont{{Pakmor}}},
  \bibnamefont{et~al.}, \bibinfo{journal}{\aap} \textbf{\bibinfo{volume}{554}},
  \bibinfo{eid}{A67} (\bibinfo{year}{2013}), \eprint{1304.2777}.

\bibitem[{\citenamefont{{Gilfanov} and {Bogd{\'a}n}}(2010)}]{gilfanov2010a}
\bibinfo{author}{\bibfnamefont{M.}~\bibnamefont{{Gilfanov}}} \bibnamefont{and}
  \bibinfo{author}{\bibfnamefont{{\'A}.}~\bibnamefont{{Bogd{\'a}n}}},
  \bibinfo{journal}{\nat} \textbf{\bibinfo{volume}{463}}, \bibinfo{pages}{924}
  (\bibinfo{year}{2010}), \eprint{1002.3359}.

\bibitem[{\citenamefont{{Badenes} et~al.}(2007)\citenamefont{{Badenes},
  {Hughes}, {Bravo}, and {Langer}}}]{badenes2007a}
\bibinfo{author}{\bibfnamefont{C.}~\bibnamefont{{Badenes}}},
  \bibinfo{author}{\bibfnamefont{J.~P.} \bibnamefont{{Hughes}}},
  \bibinfo{author}{\bibfnamefont{E.}~\bibnamefont{{Bravo}}}, \bibnamefont{and}
  \bibinfo{author}{\bibfnamefont{N.}~\bibnamefont{{Langer}}},
  \bibinfo{journal}{\apj} \textbf{\bibinfo{volume}{662}}, \bibinfo{pages}{472}
  (\bibinfo{year}{2007}), \eprint{arXiv:astro-ph/0703321}.

\bibitem[{\citenamefont{{Yamaguchi} et~al.}(2015)\citenamefont{{Yamaguchi},
  {Badenes}, {Foster}, {Bravo}, {Williams}, {Maeda}, {Nobukawa}, {Eriksen},
  {Brickhouse}, {Petre} et~al.}}]{yamaguchi2015a}
\bibinfo{author}{\bibfnamefont{H.}~\bibnamefont{{Yamaguchi}}},
  \bibinfo{author}{\bibfnamefont{C.}~\bibnamefont{{Badenes}}},
  \bibinfo{author}{\bibfnamefont{A.~R.} \bibnamefont{{Foster}}},
  \bibinfo{author}{\bibfnamefont{E.}~\bibnamefont{{Bravo}}},
  \bibinfo{author}{\bibfnamefont{B.~J.} \bibnamefont{{Williams}}},
  \bibinfo{author}{\bibfnamefont{K.}~\bibnamefont{{Maeda}}},
  \bibinfo{author}{\bibfnamefont{M.}~\bibnamefont{{Nobukawa}}},
  \bibinfo{author}{\bibfnamefont{K.~A.} \bibnamefont{{Eriksen}}},
  \bibinfo{author}{\bibfnamefont{N.~S.} \bibnamefont{{Brickhouse}}},
  \bibinfo{author}{\bibfnamefont{R.}~\bibnamefont{{Petre}}},
  \bibnamefont{et~al.}, \bibinfo{journal}{\apjl}
  \textbf{\bibinfo{volume}{801}}, \bibinfo{eid}{L31} (\bibinfo{year}{2015}),
  \eprint{1502.04255}.

\bibitem[{\citenamefont{{Ruiz-Lapuente}
  et~al.}(2004)\citenamefont{{Ruiz-Lapuente}, {Comeron}, {M{\'e}ndez}, {Canal},
  {Smartt}, {Filippenko}, {Kurucz}, {Chornock}, {Foley}, {Stanishev}
  et~al.}}]{ruiz-lapuente2004a}
\bibinfo{author}{\bibfnamefont{P.}~\bibnamefont{{Ruiz-Lapuente}}},
  \bibinfo{author}{\bibfnamefont{F.}~\bibnamefont{{Comeron}}},
  \bibinfo{author}{\bibfnamefont{J.}~\bibnamefont{{M{\'e}ndez}}},
  \bibinfo{author}{\bibfnamefont{R.}~\bibnamefont{{Canal}}},
  \bibinfo{author}{\bibfnamefont{S.~J.} \bibnamefont{{Smartt}}},
  \bibinfo{author}{\bibfnamefont{A.~V.} \bibnamefont{{Filippenko}}},
  \bibinfo{author}{\bibfnamefont{R.~L.} \bibnamefont{{Kurucz}}},
  \bibinfo{author}{\bibfnamefont{R.}~\bibnamefont{{Chornock}}},
  \bibinfo{author}{\bibfnamefont{R.~J.} \bibnamefont{{Foley}}},
  \bibinfo{author}{\bibfnamefont{V.}~\bibnamefont{{Stanishev}}},
  \bibnamefont{et~al.}, \bibinfo{journal}{\nat} \textbf{\bibinfo{volume}{431}},
  \bibinfo{pages}{1069} (\bibinfo{year}{2004}),
  \eprint{arXiv:astro-ph/0410673}.

\bibitem[{\citenamefont{{Kerzendorf} et~al.}(2009)\citenamefont{{Kerzendorf},
  {Schmidt}, {Asplund}, {Nomoto}, {Podsiadlowski}, {Frebel}, {Fesen}, and
  {Yong}}}]{kerzendorf2009a}
\bibinfo{author}{\bibfnamefont{W.~E.} \bibnamefont{{Kerzendorf}}},
  \bibinfo{author}{\bibfnamefont{B.~P.} \bibnamefont{{Schmidt}}},
  \bibinfo{author}{\bibfnamefont{M.}~\bibnamefont{{Asplund}}},
  \bibinfo{author}{\bibfnamefont{K.}~\bibnamefont{{Nomoto}}},
  \bibinfo{author}{\bibfnamefont{P.}~\bibnamefont{{Podsiadlowski}}},
  \bibinfo{author}{\bibfnamefont{A.}~\bibnamefont{{Frebel}}},
  \bibinfo{author}{\bibfnamefont{R.~A.} \bibnamefont{{Fesen}}},
  \bibnamefont{and} \bibinfo{author}{\bibfnamefont{D.}~\bibnamefont{{Yong}}},
  \bibinfo{journal}{\apj} \textbf{\bibinfo{volume}{701}}, \bibinfo{pages}{1665}
  (\bibinfo{year}{2009}), \eprint{0906.0982}.

\bibitem[{\citenamefont{{Schaefer} and {Pagnotta}}(2012)}]{schaefer2012a}
\bibinfo{author}{\bibfnamefont{B.~E.} \bibnamefont{{Schaefer}}}
  \bibnamefont{and}
  \bibinfo{author}{\bibfnamefont{A.}~\bibnamefont{{Pagnotta}}},
  \bibinfo{journal}{\nat} \textbf{\bibinfo{volume}{481}}, \bibinfo{pages}{164}
  (\bibinfo{year}{2012}).

\bibitem[{\citenamefont{{Horesh} et~al.}(2012)\citenamefont{{Horesh},
  {Kulkarni}, {Fox}, {Carpenter}, {Kasliwal}, {Ofek}, {Quimby}, {Gal-Yam},
  {Cenko}, {de Bruyn} et~al.}}]{horesh2012a}
\bibinfo{author}{\bibfnamefont{A.}~\bibnamefont{{Horesh}}},
  \bibinfo{author}{\bibfnamefont{S.~R.} \bibnamefont{{Kulkarni}}},
  \bibinfo{author}{\bibfnamefont{D.~B.} \bibnamefont{{Fox}}},
  \bibinfo{author}{\bibfnamefont{J.}~\bibnamefont{{Carpenter}}},
  \bibinfo{author}{\bibfnamefont{M.~M.} \bibnamefont{{Kasliwal}}},
  \bibinfo{author}{\bibfnamefont{E.~O.} \bibnamefont{{Ofek}}},
  \bibinfo{author}{\bibfnamefont{R.}~\bibnamefont{{Quimby}}},
  \bibinfo{author}{\bibfnamefont{A.}~\bibnamefont{{Gal-Yam}}},
  \bibinfo{author}{\bibfnamefont{S.~B.} \bibnamefont{{Cenko}}},
  \bibinfo{author}{\bibfnamefont{A.~G.} \bibnamefont{{de Bruyn}}},
  \bibnamefont{et~al.}, \bibinfo{journal}{\apj} \textbf{\bibinfo{volume}{746}},
  \bibinfo{eid}{21} (\bibinfo{year}{2012}), \eprint{1109.2912}.

\bibitem[{\citenamefont{{Chomiuk} et~al.}(2012)\citenamefont{{Chomiuk},
  {Soderberg}, {Moe}, {Chevalier}, {Rupen}, {Badenes}, {Margutti}, {Fransson},
  {Fong}, and {Dittmann}}}]{chomiuk2012a}
\bibinfo{author}{\bibfnamefont{L.}~\bibnamefont{{Chomiuk}}},
  \bibinfo{author}{\bibfnamefont{A.~M.} \bibnamefont{{Soderberg}}},
  \bibinfo{author}{\bibfnamefont{M.}~\bibnamefont{{Moe}}},
  \bibinfo{author}{\bibfnamefont{R.~A.} \bibnamefont{{Chevalier}}},
  \bibinfo{author}{\bibfnamefont{M.~P.} \bibnamefont{{Rupen}}},
  \bibinfo{author}{\bibfnamefont{C.}~\bibnamefont{{Badenes}}},
  \bibinfo{author}{\bibfnamefont{R.}~\bibnamefont{{Margutti}}},
  \bibinfo{author}{\bibfnamefont{C.}~\bibnamefont{{Fransson}}},
  \bibinfo{author}{\bibfnamefont{W.-f.} \bibnamefont{{Fong}}},
  \bibnamefont{and} \bibinfo{author}{\bibfnamefont{J.~A.}
  \bibnamefont{{Dittmann}}}, \bibinfo{journal}{\apj}
  \textbf{\bibinfo{volume}{750}}, \bibinfo{eid}{164} (\bibinfo{year}{2012}),
  \eprint{1201.0994}.

\bibitem[{\citenamefont{{Yungelson} and {Livio}}(2000)}]{yungelson2000a}
\bibinfo{author}{\bibfnamefont{L.~R.} \bibnamefont{{Yungelson}}}
  \bibnamefont{and} \bibinfo{author}{\bibfnamefont{M.}~\bibnamefont{{Livio}}},
  \bibinfo{journal}{\apj} \textbf{\bibinfo{volume}{528}}, \bibinfo{pages}{108}
  (\bibinfo{year}{2000}), \eprint{arXiv:astro-ph/9907359}.

\bibitem[{\citenamefont{{Ruiter} et~al.}(2009)\citenamefont{{Ruiter},
  {Belczynski}, and {Fryer}}}]{ruiter2009a}
\bibinfo{author}{\bibfnamefont{A.~J.} \bibnamefont{{Ruiter}}},
  \bibinfo{author}{\bibfnamefont{K.}~\bibnamefont{{Belczynski}}},
  \bibnamefont{and} \bibinfo{author}{\bibfnamefont{C.}~\bibnamefont{{Fryer}}},
  \bibinfo{journal}{\apj} \textbf{\bibinfo{volume}{699}}, \bibinfo{pages}{2026}
  (\bibinfo{year}{2009}), \eprint{0904.3108}.

\bibitem[{\citenamefont{{Ruiter} et~al.}(2011)\citenamefont{{Ruiter},
  {Belczynski}, {Sim}, {Hillebrandt}, {Fryer}, {Fink}, and
  {Kromer}}}]{ruiter2011a}
\bibinfo{author}{\bibfnamefont{A.~J.} \bibnamefont{{Ruiter}}},
  \bibinfo{author}{\bibfnamefont{K.}~\bibnamefont{{Belczynski}}},
  \bibinfo{author}{\bibfnamefont{S.~A.} \bibnamefont{{Sim}}},
  \bibinfo{author}{\bibfnamefont{W.}~\bibnamefont{{Hillebrandt}}},
  \bibinfo{author}{\bibfnamefont{C.~L.} \bibnamefont{{Fryer}}},
  \bibinfo{author}{\bibfnamefont{M.}~\bibnamefont{{Fink}}}, \bibnamefont{and}
  \bibinfo{author}{\bibfnamefont{M.}~\bibnamefont{{Kromer}}},
  \bibinfo{journal}{\mnras} \textbf{\bibinfo{volume}{417}},
  \bibinfo{pages}{408} (\bibinfo{year}{2011}), \eprint{1011.1407}.

\bibitem[{\citenamefont{{Toonen} et~al.}(2012)\citenamefont{{Toonen},
  {Nelemans}, and {Portegies Zwart}}}]{toonen2012a}
\bibinfo{author}{\bibfnamefont{S.}~\bibnamefont{{Toonen}}},
  \bibinfo{author}{\bibfnamefont{G.}~\bibnamefont{{Nelemans}}},
  \bibnamefont{and} \bibinfo{author}{\bibfnamefont{S.}~\bibnamefont{{Portegies
  Zwart}}}, \bibinfo{journal}{\aap} \textbf{\bibinfo{volume}{546}},
  \bibinfo{eid}{A70} (\bibinfo{year}{2012}), \eprint{1208.6446}.

\bibitem[{\citenamefont{{Ruiter} et~al.}(2013)\citenamefont{{Ruiter}, {Sim},
  {Pakmor}, {Kromer}, {Seitenzahl}, {Belczynski}, {Fink}, {Herzog},
  {Hillebrandt}, {R{\"o}pke} et~al.}}]{ruiter2013a}
\bibinfo{author}{\bibfnamefont{A.~J.} \bibnamefont{{Ruiter}}},
  \bibinfo{author}{\bibfnamefont{S.~A.} \bibnamefont{{Sim}}},
  \bibinfo{author}{\bibfnamefont{R.}~\bibnamefont{{Pakmor}}},
  \bibinfo{author}{\bibfnamefont{M.}~\bibnamefont{{Kromer}}},
  \bibinfo{author}{\bibfnamefont{I.~R.} \bibnamefont{{Seitenzahl}}},
  \bibinfo{author}{\bibfnamefont{K.}~\bibnamefont{{Belczynski}}},
  \bibinfo{author}{\bibfnamefont{M.}~\bibnamefont{{Fink}}},
  \bibinfo{author}{\bibfnamefont{M.}~\bibnamefont{{Herzog}}},
  \bibinfo{author}{\bibfnamefont{W.}~\bibnamefont{{Hillebrandt}}},
  \bibinfo{author}{\bibfnamefont{F.~K.} \bibnamefont{{R{\"o}pke}}},
  \bibnamefont{et~al.}, \bibinfo{journal}{\mnras}
  \textbf{\bibinfo{volume}{429}}, \bibinfo{pages}{1425} (\bibinfo{year}{2013}),
  \eprint{1209.0645}.

\bibitem[{\citenamefont{{Mennekens} et~al.}(2010)\citenamefont{{Mennekens},
  {Vanbeveren}, {De Greve}, and {De Donder}}}]{mennekens2010a}
\bibinfo{author}{\bibfnamefont{N.}~\bibnamefont{{Mennekens}}},
  \bibinfo{author}{\bibfnamefont{D.}~\bibnamefont{{Vanbeveren}}},
  \bibinfo{author}{\bibfnamefont{J.~P.} \bibnamefont{{De Greve}}},
  \bibnamefont{and} \bibinfo{author}{\bibfnamefont{E.}~\bibnamefont{{De
  Donder}}}, \bibinfo{journal}{\aap} \textbf{\bibinfo{volume}{515}},
  \bibinfo{eid}{A89} (\bibinfo{year}{2010}), \eprint{1003.2491}.

\bibitem[{\citenamefont{{Wang} et~al.}(2012)\citenamefont{{Wang}, {Wang},
  {Filippenko}, {Aldering}, {Antilogus}, {Arnett}, {Baade}, {Baron}, {Barris},
  {Benetti} et~al.}}]{wang2012a}
\bibinfo{author}{\bibfnamefont{X.}~\bibnamefont{{Wang}}},
  \bibinfo{author}{\bibfnamefont{L.}~\bibnamefont{{Wang}}},
  \bibinfo{author}{\bibfnamefont{A.~V.} \bibnamefont{{Filippenko}}},
  \bibinfo{author}{\bibfnamefont{G.}~\bibnamefont{{Aldering}}},
  \bibinfo{author}{\bibfnamefont{P.}~\bibnamefont{{Antilogus}}},
  \bibinfo{author}{\bibfnamefont{D.}~\bibnamefont{{Arnett}}},
  \bibinfo{author}{\bibfnamefont{D.}~\bibnamefont{{Baade}}},
  \bibinfo{author}{\bibfnamefont{E.}~\bibnamefont{{Baron}}},
  \bibinfo{author}{\bibfnamefont{B.~J.} \bibnamefont{{Barris}}},
  \bibinfo{author}{\bibfnamefont{S.}~\bibnamefont{{Benetti}}},
  \bibnamefont{et~al.}, \bibinfo{journal}{\apj} \textbf{\bibinfo{volume}{749}},
  \bibinfo{eid}{126} (\bibinfo{year}{2012}), \eprint{1110.5809},
  \urlprefix\url{http://adsabs.harvard.edu/abs/2012ApJ...749..126W}.

\bibitem[{\citenamefont{{Badenes} and {Maoz}}(2012)}]{badenes2012a}
\bibinfo{author}{\bibfnamefont{C.}~\bibnamefont{{Badenes}}} \bibnamefont{and}
  \bibinfo{author}{\bibfnamefont{D.}~\bibnamefont{{Maoz}}},
  \bibinfo{journal}{\apjl} \textbf{\bibinfo{volume}{749}}, \bibinfo{eid}{L11}
  (\bibinfo{year}{2012}), \eprint{1202.5472}.

\bibitem[{\citenamefont{{Seitenzahl}
  et~al.}(2013{\natexlab{a}})\citenamefont{{Seitenzahl}, {Ciaraldi-Schoolmann},
  {R{\"o}pke}, {Fink}, {Hillebrandt}, {Kromer}, {Pakmor}, {Ruiter}, {Sim}, and
  {Taubenberger}}}]{seitenzahl2013a}
\bibinfo{author}{\bibfnamefont{I.~R.} \bibnamefont{{Seitenzahl}}},
  \bibinfo{author}{\bibfnamefont{F.}~\bibnamefont{{Ciaraldi-Schoolmann}}},
  \bibinfo{author}{\bibfnamefont{F.~K.} \bibnamefont{{R{\"o}pke}}},
  \bibinfo{author}{\bibfnamefont{M.}~\bibnamefont{{Fink}}},
  \bibinfo{author}{\bibfnamefont{W.}~\bibnamefont{{Hillebrandt}}},
  \bibinfo{author}{\bibfnamefont{M.}~\bibnamefont{{Kromer}}},
  \bibinfo{author}{\bibfnamefont{R.}~\bibnamefont{{Pakmor}}},
  \bibinfo{author}{\bibfnamefont{A.~J.} \bibnamefont{{Ruiter}}},
  \bibinfo{author}{\bibfnamefont{S.~A.} \bibnamefont{{Sim}}}, \bibnamefont{and}
  \bibinfo{author}{\bibfnamefont{S.}~\bibnamefont{{Taubenberger}}},
  \bibinfo{journal}{\mnras} \textbf{\bibinfo{volume}{429}},
  \bibinfo{pages}{1156} (\bibinfo{year}{2013}{\natexlab{a}}),
  \eprint{1211.3015}.

\bibitem[{\citenamefont{{Hillebrandt} and {Niemeyer}}(2000)}]{hillebrandt2000a}
\bibinfo{author}{\bibfnamefont{W.}~\bibnamefont{{Hillebrandt}}}
  \bibnamefont{and} \bibinfo{author}{\bibfnamefont{J.~C.}
  \bibnamefont{{Niemeyer}}}, \bibinfo{journal}{\araa}
  \textbf{\bibinfo{volume}{38}}, \bibinfo{pages}{191} (\bibinfo{year}{2000}),
  \eprint{arXiv:astro-ph/0006305}.

\bibitem[{\citenamefont{{Liu} et~al.}(2013)\citenamefont{{Liu}, {Pakmor},
  {Seitenzahl}, {Hillebrandt}, {Kromer}, {R{\"o}pke}, {Edelmann},
  {Taubenberger}, {Maeda}, {Wang} et~al.}}]{liu2013a}
\bibinfo{author}{\bibfnamefont{Z.-W.} \bibnamefont{{Liu}}},
  \bibinfo{author}{\bibfnamefont{R.}~\bibnamefont{{Pakmor}}},
  \bibinfo{author}{\bibfnamefont{I.~R.} \bibnamefont{{Seitenzahl}}},
  \bibinfo{author}{\bibfnamefont{W.}~\bibnamefont{{Hillebrandt}}},
  \bibinfo{author}{\bibfnamefont{M.}~\bibnamefont{{Kromer}}},
  \bibinfo{author}{\bibfnamefont{F.~K.} \bibnamefont{{R{\"o}pke}}},
  \bibinfo{author}{\bibfnamefont{P.}~\bibnamefont{{Edelmann}}},
  \bibinfo{author}{\bibfnamefont{S.}~\bibnamefont{{Taubenberger}}},
  \bibinfo{author}{\bibfnamefont{K.}~\bibnamefont{{Maeda}}},
  \bibinfo{author}{\bibfnamefont{B.}~\bibnamefont{{Wang}}},
  \bibnamefont{et~al.}, \bibinfo{journal}{\apj} \textbf{\bibinfo{volume}{774}},
  \bibinfo{eid}{37} (\bibinfo{year}{2013}), \eprint{1307.5579}.

\bibitem[{\citenamefont{{Seitenzahl}
  et~al.}(2013{\natexlab{b}})\citenamefont{{Seitenzahl}, {Cescutti},
  {R{\"o}pke}, {Ruiter}, and {Pakmor}}}]{seitenzahl2013b}
\bibinfo{author}{\bibfnamefont{I.~R.} \bibnamefont{{Seitenzahl}}},
  \bibinfo{author}{\bibfnamefont{G.}~\bibnamefont{{Cescutti}}},
  \bibinfo{author}{\bibfnamefont{F.~K.} \bibnamefont{{R{\"o}pke}}},
  \bibinfo{author}{\bibfnamefont{A.~J.} \bibnamefont{{Ruiter}}},
  \bibnamefont{and} \bibinfo{author}{\bibfnamefont{R.}~\bibnamefont{{Pakmor}}},
  \bibinfo{journal}{\aap} \textbf{\bibinfo{volume}{559}}, \bibinfo{eid}{L5}
  (\bibinfo{year}{2013}{\natexlab{b}}), \eprint{1309.2397}.

\bibitem[{\citenamefont{{Seitenzahl} et~al.}(2015)\citenamefont{{Seitenzahl},
  {Summa}, {Krau{\ss}}, {Sim}, {Diehl}, {Els{\"a}sser}, {Fink}, {Hillebrandt},
  {Kromer}, {Maeda} et~al.}}]{seitenzahl2015a}
\bibinfo{author}{\bibfnamefont{I.~R.} \bibnamefont{{Seitenzahl}}},
  \bibinfo{author}{\bibfnamefont{A.}~\bibnamefont{{Summa}}},
  \bibinfo{author}{\bibfnamefont{F.}~\bibnamefont{{Krau{\ss}}}},
  \bibinfo{author}{\bibfnamefont{S.~A.} \bibnamefont{{Sim}}},
  \bibinfo{author}{\bibfnamefont{R.}~\bibnamefont{{Diehl}}},
  \bibinfo{author}{\bibfnamefont{D.}~\bibnamefont{{Els{\"a}sser}}},
  \bibinfo{author}{\bibfnamefont{M.}~\bibnamefont{{Fink}}},
  \bibinfo{author}{\bibfnamefont{W.}~\bibnamefont{{Hillebrandt}}},
  \bibinfo{author}{\bibfnamefont{M.}~\bibnamefont{{Kromer}}},
  \bibinfo{author}{\bibfnamefont{K.}~\bibnamefont{{Maeda}}},
  \bibnamefont{et~al.}, \bibinfo{journal}{\mnras}
  \textbf{\bibinfo{volume}{447}}, \bibinfo{pages}{1484} (\bibinfo{year}{2015}),
  \eprint{1412.0835}.

\bibitem[{\citenamefont{{Timmes} and {Woosley}}(1992)}]{timmes1992a}
\bibinfo{author}{\bibfnamefont{F.~X.} \bibnamefont{{Timmes}}} \bibnamefont{and}
  \bibinfo{author}{\bibfnamefont{S.~E.} \bibnamefont{{Woosley}}},
  \bibinfo{journal}{\apj} \textbf{\bibinfo{volume}{396}}, \bibinfo{pages}{649}
  (\bibinfo{year}{1992}).

\bibitem[{\citenamefont{{Bethe}}(1990)}]{bethe1990a}
\bibinfo{author}{\bibfnamefont{H.~A.} \bibnamefont{{Bethe}}},
  \bibinfo{journal}{Reviews of Modern Physics} \textbf{\bibinfo{volume}{62}},
  \bibinfo{pages}{801} (\bibinfo{year}{1990}).

\bibitem[{\citenamefont{{Nomoto} et~al.}(1984)\citenamefont{{Nomoto},
  {Thielemann}, and {Yokoi}}}]{nomoto1984a}
\bibinfo{author}{\bibfnamefont{K.}~\bibnamefont{{Nomoto}}},
  \bibinfo{author}{\bibfnamefont{F.-K.} \bibnamefont{{Thielemann}}},
  \bibnamefont{and} \bibinfo{author}{\bibfnamefont{K.}~\bibnamefont{{Yokoi}}},
  \bibinfo{journal}{\apj} \textbf{\bibinfo{volume}{286}}, \bibinfo{pages}{644}
  (\bibinfo{year}{1984}).

\bibitem[{\citenamefont{{Kunugise} and {Iwamoto}}(2007)}]{kunugise2007a}
\bibinfo{author}{\bibfnamefont{T.}~\bibnamefont{{Kunugise}}} \bibnamefont{and}
  \bibinfo{author}{\bibfnamefont{K.}~\bibnamefont{{Iwamoto}}},
  \bibinfo{journal}{\pasj} \textbf{\bibinfo{volume}{59}}, \bibinfo{pages}{L57+}
  (\bibinfo{year}{2007}).

\bibitem[{\citenamefont{{Odrzywolek} and {Plewa}}(2011)}]{odrzywolek2011a}
\bibinfo{author}{\bibfnamefont{A.}~\bibnamefont{{Odrzywolek}}}
  \bibnamefont{and} \bibinfo{author}{\bibfnamefont{T.}~\bibnamefont{{Plewa}}},
  \bibinfo{journal}{\aap} \textbf{\bibinfo{volume}{529}}, \bibinfo{eid}{A156}
  (\bibinfo{year}{2011}), \eprint{1006.0490}.

\bibitem[{\citenamefont{{Plewa}}(2007)}]{plewa2007a}
\bibinfo{author}{\bibfnamefont{T.}~\bibnamefont{{Plewa}}},
  \bibinfo{journal}{\apj} \textbf{\bibinfo{volume}{657}}, \bibinfo{pages}{942}
  (\bibinfo{year}{2007}), \eprint{arXiv:astro-ph/0611776}.

\bibitem[{\citenamefont{{Khokhlov}}(1991)}]{khokhlov1991a}
\bibinfo{author}{\bibfnamefont{A.~M.} \bibnamefont{{Khokhlov}}},
  \bibinfo{journal}{\aap} \textbf{\bibinfo{volume}{245}}, \bibinfo{pages}{114}
  (\bibinfo{year}{1991}).

\bibitem[{\citenamefont{{Plewa} et~al.}(2004)\citenamefont{{Plewa}, {Calder},
  and {Lamb}}}]{plewa2004a}
\bibinfo{author}{\bibfnamefont{T.}~\bibnamefont{{Plewa}}},
  \bibinfo{author}{\bibfnamefont{A.~C.} \bibnamefont{{Calder}}},
  \bibnamefont{and} \bibinfo{author}{\bibfnamefont{D.~Q.}
  \bibnamefont{{Lamb}}}, \bibinfo{journal}{\apjl}
  \textbf{\bibinfo{volume}{612}}, \bibinfo{pages}{L37} (\bibinfo{year}{2004}),
  \eprint{arXiv:astro-ph/0405163}.

\bibitem[{\citenamefont{{Seitenzahl}
  et~al.}(2009{\natexlab{b}})\citenamefont{{Seitenzahl}, {Meakin}, {Lamb}, and
  {Truran}}}]{seitenzahl2009c}
\bibinfo{author}{\bibfnamefont{I.~R.} \bibnamefont{{Seitenzahl}}},
  \bibinfo{author}{\bibfnamefont{C.~A.} \bibnamefont{{Meakin}}},
  \bibinfo{author}{\bibfnamefont{D.~Q.} \bibnamefont{{Lamb}}},
  \bibnamefont{and} \bibinfo{author}{\bibfnamefont{J.~W.}
  \bibnamefont{{Truran}}}, \bibinfo{journal}{\apj}
  \textbf{\bibinfo{volume}{700}}, \bibinfo{pages}{642}
  (\bibinfo{year}{2009}{\natexlab{b}}), \eprint{0905.3104}.

\bibitem[{\citenamefont{{Hillebrandt} et~al.}(2013)\citenamefont{{Hillebrandt},
  {Kromer}, {R{\"o}pke}, and {Ruiter}}}]{hillebrandt2013a}
\bibinfo{author}{\bibfnamefont{W.}~\bibnamefont{{Hillebrandt}}},
  \bibinfo{author}{\bibfnamefont{M.}~\bibnamefont{{Kromer}}},
  \bibinfo{author}{\bibfnamefont{F.~K.} \bibnamefont{{R{\"o}pke}}},
  \bibnamefont{and} \bibinfo{author}{\bibfnamefont{A.~J.}
  \bibnamefont{{Ruiter}}}, \bibinfo{journal}{Frontiers of Physics}
  \textbf{\bibinfo{volume}{8}}, \bibinfo{pages}{116} (\bibinfo{year}{2013}),
  \eprint{1302.6420}.

\bibitem[{\citenamefont{{Ciaraldi-Schoolmann}
  et~al.}(2013)\citenamefont{{Ciaraldi-Schoolmann}, {Seitenzahl}, and
  {R{\"o}pke}}}]{ciaraldi2013a}
\bibinfo{author}{\bibfnamefont{F.}~\bibnamefont{{Ciaraldi-Schoolmann}}},
  \bibinfo{author}{\bibfnamefont{I.~R.} \bibnamefont{{Seitenzahl}}},
  \bibnamefont{and} \bibinfo{author}{\bibfnamefont{F.~K.}
  \bibnamefont{{R{\"o}pke}}}, \bibinfo{journal}{\aap}
  \textbf{\bibinfo{volume}{559}}, \bibinfo{eid}{A117} (\bibinfo{year}{2013}),
  \eprint{1307.8146}.

\bibitem[{\citenamefont{{Shapiro} and {Teukolsky}}(1983)}]{shapiro1983a}
\bibinfo{author}{\bibfnamefont{S.~L.} \bibnamefont{{Shapiro}}}
  \bibnamefont{and} \bibinfo{author}{\bibfnamefont{S.~A.}
  \bibnamefont{{Teukolsky}}}, \emph{\bibinfo{title}{Black Holes, White Dwarfs,
  and Neutron Stars}} (\bibinfo{publisher}{John Wiley \& Sons},
  \bibinfo{address}{New York}, \bibinfo{year}{1983}).

\bibitem[{\citenamefont{{Woosley}}(1997)}]{woosley1997b}
\bibinfo{author}{\bibfnamefont{S.~E.} \bibnamefont{{Woosley}}},
  \bibinfo{journal}{\apj} \textbf{\bibinfo{volume}{476}}, \bibinfo{pages}{801}
  (\bibinfo{year}{1997}).

\bibitem[{\citenamefont{{Langanke} and
  {Mart{\'{\i}}nez-Pinedo}}(2001)}]{langanke2001a}
\bibinfo{author}{\bibfnamefont{K.}~\bibnamefont{{Langanke}}} \bibnamefont{and}
  \bibinfo{author}{\bibfnamefont{G.}~\bibnamefont{{Mart{\'{\i}}nez-Pinedo}}},
  \bibinfo{journal}{Atomic Data and Nuclear Data Tables}
  \textbf{\bibinfo{volume}{79}}, \bibinfo{pages}{1} (\bibinfo{year}{2001}).

\bibitem[{\citenamefont{{Seitenzahl}
  et~al.}(2009{\natexlab{c}})\citenamefont{{Seitenzahl}, {Townsley}, {Peng},
  and {Truran}}}]{seitenzahl2009a}
\bibinfo{author}{\bibfnamefont{I.~R.} \bibnamefont{{Seitenzahl}}},
  \bibinfo{author}{\bibfnamefont{D.~M.} \bibnamefont{{Townsley}}},
  \bibinfo{author}{\bibfnamefont{F.}~\bibnamefont{{Peng}}}, \bibnamefont{and}
  \bibinfo{author}{\bibfnamefont{J.~W.} \bibnamefont{{Truran}}},
  \bibinfo{journal}{Atomic Data and Nuclear Data Tables}
  \textbf{\bibinfo{volume}{95}}, \bibinfo{pages}{96}
  (\bibinfo{year}{2009}{\natexlab{c}}).

\bibitem[{\citenamefont{{Itoh} et~al.}(1996)\citenamefont{{Itoh}, {Hayashi},
  {Nishikawa}, and {Kohyama}}}]{itoh1996a}
\bibinfo{author}{\bibfnamefont{N.}~\bibnamefont{{Itoh}}},
  \bibinfo{author}{\bibfnamefont{H.}~\bibnamefont{{Hayashi}}},
  \bibinfo{author}{\bibfnamefont{A.}~\bibnamefont{{Nishikawa}}},
  \bibnamefont{and}
  \bibinfo{author}{\bibfnamefont{Y.}~\bibnamefont{{Kohyama}}},
  \bibinfo{journal}{\apjs} \textbf{\bibinfo{volume}{102}}, \bibinfo{pages}{411}
  (\bibinfo{year}{1996}).

\bibitem[{\citenamefont{{Consortium} et~al.}(2013)\citenamefont{{Consortium},
  {:}, {Seoane}, {Aoudia}, {Audley}, {Auger}, {Babak}, {Baker}, {Barausse},
  {Barke} et~al.}}]{elisa2013a}
\bibinfo{author}{\bibfnamefont{T.~e.} \bibnamefont{{Consortium}}},
  \bibinfo{author}{\bibnamefont{{:}}}, \bibinfo{author}{\bibfnamefont{P.~A.}
  \bibnamefont{{Seoane}}},
  \bibinfo{author}{\bibfnamefont{S.}~\bibnamefont{{Aoudia}}},
  \bibinfo{author}{\bibfnamefont{H.}~\bibnamefont{{Audley}}},
  \bibinfo{author}{\bibfnamefont{G.}~\bibnamefont{{Auger}}},
  \bibinfo{author}{\bibfnamefont{S.}~\bibnamefont{{Babak}}},
  \bibinfo{author}{\bibfnamefont{J.}~\bibnamefont{{Baker}}},
  \bibinfo{author}{\bibfnamefont{E.}~\bibnamefont{{Barausse}}},
  \bibinfo{author}{\bibfnamefont{S.}~\bibnamefont{{Barke}}},
  \bibnamefont{et~al.}, \bibinfo{journal}{ArXiv e-prints}
  (\bibinfo{year}{2013}), \eprint{1305.5720}.

\bibitem[{\citenamefont{{Nelemans} et~al.}(2001)\citenamefont{{Nelemans},
  {Yungelson}, and {Portegies Zwart}}}]{nelemans2001a}
\bibinfo{author}{\bibfnamefont{G.}~\bibnamefont{{Nelemans}}},
  \bibinfo{author}{\bibfnamefont{L.~R.} \bibnamefont{{Yungelson}}},
  \bibnamefont{and} \bibinfo{author}{\bibfnamefont{S.~F.}
  \bibnamefont{{Portegies Zwart}}}, \bibinfo{journal}{\aap}
  \textbf{\bibinfo{volume}{375}}, \bibinfo{pages}{890} (\bibinfo{year}{2001}),
  \eprint{astro-ph/0105221}.

\bibitem[{\citenamefont{{Ruiter} et~al.}(2010)\citenamefont{{Ruiter},
  {Belczynski}, {Sim}, {Hillebrandt}, {Fryer}, {Fink}, and
  {Kromer}}}]{ruiter2010a}
\bibinfo{author}{\bibfnamefont{A.~J.} \bibnamefont{{Ruiter}}},
  \bibinfo{author}{\bibfnamefont{K.}~\bibnamefont{{Belczynski}}},
  \bibinfo{author}{\bibfnamefont{S.~A.} \bibnamefont{{Sim}}},
  \bibinfo{author}{\bibfnamefont{W.}~\bibnamefont{{Hillebrandt}}},
  \bibinfo{author}{\bibfnamefont{C.~L.} \bibnamefont{{Fryer}}},
  \bibinfo{author}{\bibfnamefont{M.}~\bibnamefont{{Fink}}}, \bibnamefont{and}
  \bibinfo{author}{\bibfnamefont{M.}~\bibnamefont{{Kromer}}}
  (\bibinfo{year}{2010}), \eprint{1011.1407}.

\bibitem[{\citenamefont{{Dan} et~al.}(2011)\citenamefont{{Dan}, {Rosswog},
  {Guillochon}, and {Ramirez-Ruiz}}}]{dan2011a}
\bibinfo{author}{\bibfnamefont{M.}~\bibnamefont{{Dan}}},
  \bibinfo{author}{\bibfnamefont{S.}~\bibnamefont{{Rosswog}}},
  \bibinfo{author}{\bibfnamefont{J.}~\bibnamefont{{Guillochon}}},
  \bibnamefont{and}
  \bibinfo{author}{\bibfnamefont{E.}~\bibnamefont{{Ramirez-Ruiz}}},
  \bibinfo{journal}{\apj} \textbf{\bibinfo{volume}{737}}, \bibinfo{eid}{89}
  (\bibinfo{year}{2011}), \eprint{1101.5132}.

\bibitem[{\citenamefont{{Falta} et~al.}(2011)\citenamefont{{Falta}, {Fisher},
  and {Khanna}}}]{falta2011a}
\bibinfo{author}{\bibfnamefont{D.}~\bibnamefont{{Falta}}},
  \bibinfo{author}{\bibfnamefont{R.}~\bibnamefont{{Fisher}}}, \bibnamefont{and}
  \bibinfo{author}{\bibfnamefont{G.}~\bibnamefont{{Khanna}}},
  \bibinfo{journal}{Physical Review Letters} \textbf{\bibinfo{volume}{106}},
  \bibinfo{eid}{201103} (\bibinfo{year}{2011}), \eprint{1011.6387}.

\bibitem[{\citenamefont{{Falta} and {Fisher}}(2011)}]{falta2011b}
\bibinfo{author}{\bibfnamefont{D.}~\bibnamefont{{Falta}}} \bibnamefont{and}
  \bibinfo{author}{\bibfnamefont{R.}~\bibnamefont{{Fisher}}},
  \bibinfo{journal}{\prd} \textbf{\bibinfo{volume}{84}}, \bibinfo{eid}{124062}
  (\bibinfo{year}{2011}), \eprint{1112.2782}.

\bibitem[{\citenamefont{{Kawamura} et~al.}(2011)\citenamefont{{Kawamura},
  {Ando}, {Seto}, {Sato}, {Nakamura}, {Tsubono}, {Kanda}, {Tanaka}, {Yokoyama},
  {Funaki} et~al.}}]{kawamura2011a}
\bibinfo{author}{\bibfnamefont{S.}~\bibnamefont{{Kawamura}}},
  \bibinfo{author}{\bibfnamefont{M.}~\bibnamefont{{Ando}}},
  \bibinfo{author}{\bibfnamefont{N.}~\bibnamefont{{Seto}}},
  \bibinfo{author}{\bibfnamefont{S.}~\bibnamefont{{Sato}}},
  \bibinfo{author}{\bibfnamefont{T.}~\bibnamefont{{Nakamura}}},
  \bibinfo{author}{\bibfnamefont{K.}~\bibnamefont{{Tsubono}}},
  \bibinfo{author}{\bibfnamefont{N.}~\bibnamefont{{Kanda}}},
  \bibinfo{author}{\bibfnamefont{T.}~\bibnamefont{{Tanaka}}},
  \bibinfo{author}{\bibfnamefont{J.}~\bibnamefont{{Yokoyama}}},
  \bibinfo{author}{\bibfnamefont{I.}~\bibnamefont{{Funaki}}},
  \bibnamefont{et~al.}, \bibinfo{journal}{Classical and Quantum Gravity}
  \textbf{\bibinfo{volume}{28}}, \bibinfo{eid}{094011} (\bibinfo{year}{2011}).

\bibitem[{\citenamefont{{Yagi} and {Seto}}(2011)}]{yagi2011a}
\bibinfo{author}{\bibfnamefont{K.}~\bibnamefont{{Yagi}}} \bibnamefont{and}
  \bibinfo{author}{\bibfnamefont{N.}~\bibnamefont{{Seto}}},
  \bibinfo{journal}{\prd} \textbf{\bibinfo{volume}{83}}, \bibinfo{eid}{044011}
  (\bibinfo{year}{2011}), \eprint{1101.3940}.

\bibitem[{\citenamefont{{Blanchet} et~al.}(1990)\citenamefont{{Blanchet},
  {Damour}, and {Schaefer}}}]{blanchet1990a}
\bibinfo{author}{\bibfnamefont{L.}~\bibnamefont{{Blanchet}}},
  \bibinfo{author}{\bibfnamefont{T.}~\bibnamefont{{Damour}}}, \bibnamefont{and}
  \bibinfo{author}{\bibfnamefont{G.}~\bibnamefont{{Schaefer}}},
  \bibinfo{journal}{\mnras} \textbf{\bibinfo{volume}{242}},
  \bibinfo{pages}{289} (\bibinfo{year}{1990}).

\bibitem[{\citenamefont{{M{\"u}ller} and {Janka}}(1997)}]{mueller1997a}
\bibinfo{author}{\bibfnamefont{E.}~\bibnamefont{{M{\"u}ller}}}
  \bibnamefont{and} \bibinfo{author}{\bibfnamefont{H.-T.}
  \bibnamefont{{Janka}}}, \bibinfo{journal}{\aap}
  \textbf{\bibinfo{volume}{317}}, \bibinfo{pages}{140} (\bibinfo{year}{1997}).

\bibitem[{\citenamefont{{Nakamura} and {Oohara}}(1989)}]{nakamura1989a}
\bibinfo{author}{\bibfnamefont{T.}~\bibnamefont{{Nakamura}}} \bibnamefont{and}
  \bibinfo{author}{\bibfnamefont{K.-I.} \bibnamefont{{Oohara}}},
  \emph{\bibinfo{title}{{Methods in 3 D numerical relativity.}}}
  (\bibinfo{year}{1989}), pp. \bibinfo{pages}{254--280}.

\bibitem[{\citenamefont{{Moenchmeyer} et~al.}(1991)\citenamefont{{Moenchmeyer},
  {Schaefer}, {M{\"u}ller}, and {Kates}}}]{moenchmeyer1991a}
\bibinfo{author}{\bibfnamefont{R.}~\bibnamefont{{Moenchmeyer}}},
  \bibinfo{author}{\bibfnamefont{G.}~\bibnamefont{{Schaefer}}},
  \bibinfo{author}{\bibfnamefont{E.}~\bibnamefont{{M{\"u}ller}}},
  \bibnamefont{and} \bibinfo{author}{\bibfnamefont{R.~E.}
  \bibnamefont{{Kates}}}, \bibinfo{journal}{\aap}
  \textbf{\bibinfo{volume}{246}}, \bibinfo{pages}{417} (\bibinfo{year}{1991}).

\bibitem[{\citenamefont{{Misner} et~al.}(1973)\citenamefont{{Misner}, {Thorne},
  and {Wheeler}}}]{misner1973a}
\bibinfo{author}{\bibfnamefont{C.~W.} \bibnamefont{{Misner}}},
  \bibinfo{author}{\bibfnamefont{K.~S.} \bibnamefont{{Thorne}}},
  \bibnamefont{and} \bibinfo{author}{\bibfnamefont{J.~A.}
  \bibnamefont{{Wheeler}}}, \emph{\bibinfo{title}{Gravitation}}
  (\bibinfo{year}{1973}).

\bibitem[{\citenamefont{{Moore} et~al.}(2015)\citenamefont{{Moore}, {Cole}, and
  {Berry}}}]{moore2015a}
\bibinfo{author}{\bibfnamefont{C.~J.} \bibnamefont{{Moore}}},
  \bibinfo{author}{\bibfnamefont{R.~H.} \bibnamefont{{Cole}}},
  \bibnamefont{and} \bibinfo{author}{\bibfnamefont{C.~P.~L.}
  \bibnamefont{{Berry}}}, \bibinfo{journal}{Classical and Quantum Gravity}
  \textbf{\bibinfo{volume}{32}}, \bibinfo{eid}{015014} (\bibinfo{year}{2015}),
  \eprint{1408.0740}.

\bibitem[{\citenamefont{{M{\"u}ller}}(1982)}]{mueller1982b}
\bibinfo{author}{\bibfnamefont{E.}~\bibnamefont{{M{\"u}ller}}},
  \bibinfo{journal}{\aap} \textbf{\bibinfo{volume}{114}}, \bibinfo{pages}{53}
  (\bibinfo{year}{1982}).

\bibitem[{\citenamefont{{M{\"u}ller} et~al.}(2012)\citenamefont{{M{\"u}ller},
  {Janka}, and {Wongwathanarat}}}]{mueller2012b}
\bibinfo{author}{\bibfnamefont{E.}~\bibnamefont{{M{\"u}ller}}},
  \bibinfo{author}{\bibfnamefont{H.-T.} \bibnamefont{{Janka}}},
  \bibnamefont{and}
  \bibinfo{author}{\bibfnamefont{A.}~\bibnamefont{{Wongwathanarat}}},
  \bibinfo{journal}{\aap} \textbf{\bibinfo{volume}{537}}, \bibinfo{eid}{A63}
  (\bibinfo{year}{2012}), \eprint{1106.6301}.

\end{thebibliography}
\end{document}